\documentclass[%
 reprint,
 amsmath,amssymb,
 aps,
pra,
]{revtex4-2}

\usepackage{subfigure}
\subfiguretopcaptrue
\usepackage{graphicx}
\usepackage{dcolumn}
\usepackage{bm}
\usepackage{amsmath,amsfonts,amssymb}

\begin{document}


\title{Optimized pulses for population transfer via laser induced continuum structures}

\author{Dionisis Stefanatos}
\email{dionisis@post.harvard.edu}

\author{Emmanuel Paspalakis}

\affiliation{Materials Science Department, School of Natural Sciences, University of Patras, Patras 26504, Greece}

\date{\today}

\begin{abstract}
We use optimal control in order to find the optimal shapes of pulses maximizing the population transfer between two bound states which are coupled via a continuum of states. We find that the optimal bounded controls acquire the bang-interior and interior-bang form, with the bang part corresponding to the maximum allowed control value and the interior part to values between zero and the maximum. Then, we use numerical optimal control to obtain the switching times and the interior control values. We compare our results with those obtained using Gaussian STIRAP pulses, and find that the optimal method performs better, with the extent of improvement depending on the effective two-photon detuning and the size of incoherent losses.  When we consider effective two-photon resonance, the improvement is more dramatic for larger incoherent losses, while when we take into account the effective two-photon detuning, the improvement is better for smaller incoherent losses. We also obtain how the transfer efficiency increases with increasing absolute value of the Fano factor. The present work is expected to find application in areas where the population transfer between two bound states through a continuum structure plays an important role, for example coherence effects, like population trapping and electromagnetically induced transparency, optical analogs for light waves propagating in waveguide-based photonic structures, and qubits coupled via a continuum of bosonic or waveguide modes.
\end{abstract}

\maketitle

\section{Introduction}

\label{sec:intro}

The problem of efficient population transfer between two bound states through a laser induced continuum structure \cite{Knight90}, has attracted considerable attention over the years \cite{Carroll92,Nakajima94,Caroll96,Vitanov97,Yatsenko97,Unanyan98,Paspalakis97,Paspalakis98,Buffa98,Tran99,Han07,Han08,Cederbaum15}. In these works, the population transfer is accomplished using a counterintuitive STIRAP  pulse sequence \cite{Bergmann98,Kral07,Vitanov17,Bergmann19}, which in most cases consists of two delayed Gaussian pulses. The method has been experimentally demonstrated in helium atoms \cite{Peters05,Peters07}, while other interesting applications include coherence effects \cite{Thanopulos06,Thanopulos10}, like population trapping and electromagnetically induced transparency, photonics \cite{Limonov17}, specifically optical analogs for light waves propagating in waveguide-based photonic structures \cite{Longhi08,Longhi21}, and qubits coupled through a bosonic structural continuum \cite{Huang19} or the quasi-continuous spectrum of modes in a waveguide \cite{Kannan20}. There is actually a plethora of systems where the population transfer between two bound states coupled via a continuum could be useful \cite{Hsu16}. Recently, the method has been extended to the case where multilevel degenerate states interact through a common continuum structure \cite{Zlatanov21}.

In the present work, we first evaluate the performance of a simple sin-cos protocol, where the mixing angle of the applied fields varies linearly with time, and find that it performs worse than the Gaussian STIRAP protocol. We then use optimal control \cite{Bryson} to find the optimal shapes of pulses which maximize the population transfer between two bound states coupled through a continuum structure. We consider bounded controls and using an elementary theoretical analysis we explain that the optimal pulses have the bang-interior and interior-bang form, where the bang part corresponds to the maximum allowed control value, while the interior part corresponds to control values between zero and this maximum bound. The analytic determination of the switching times as well as of the interior control values is a formidable task, thus we recourse to numerical optimal control \cite{bocop}. We are benchmarking our method by applying it to the system used in Ref. \cite{Vitanov97}, and compare our results with those obtained there using Gaussian STIRAP pulses. We find that the optimal method outperforms STIRAP pulses. The extent of improvement depends on the effective two-photon detuning and the size of incoherent losses.  Specifically, in the case of effective two-photon resonance, the improvement is more dramatic for larger incoherent losses, while when the effective two-photon detuning is taken into account, the improvement is better for smaller incoherent losses. We also demonstrate the increase in transfer efficiency with the increase in the absolute value of the Fano factor. Note that numerical optimal control has been previously used to increase the population transfer between two bound states coupled via a continuum \cite{Buffa98}. The main difference of the present approach is that we use as control variables only the envelopes of ionization pulses, while in Ref. \cite{Buffa98} the effective two-photon detuning was also used as an extra control variable.

The structure of the paper is as follows. In the next section we formulate the problem and summarize the findings of Ref. \cite{Vitanov97} with Gaussian pulses, while in section \ref{sec:linear} we evaluate the performance of the simple sin-cos protocol. In section \ref{optimal_solution} we analyze the optimal control problem, while in section \ref{sec:results}
we present the results of numerical optimization and compare them with those of Ref. \cite{Vitanov97}. Section \ref{sec:conclusion} concludes this work.

\section{Population transfer through continuum states}

\label{sec:formulation}

The dynamics of two bound states coupled by two laser pulses through a continuum of intermediate states is governed by the following equation \cite{Vitanov97}
\begin{equation}
\label{LICS}
i
\left[
\begin{array}{c}
\dot{c}_g\\
\\
\dot{c}_e
\end{array}
\right]
=
\left[
\begin{array}{cc}
\Sigma_g-\frac{1}{2}i\Gamma_g & -\frac{1}{2}\sqrt{\Gamma^p_g\Gamma^{s}_e}(q+i)\\
& \\
-\frac{1}{2}\sqrt{\Gamma^p_g\Gamma^s_e}(q+i) & \Sigma_e-\frac{1}{2}i\Gamma_e+D
\end{array}
\right]
\left[
\begin{array}{c}
c_g\\
\\
c_e
\end{array}
\right],
\end{equation}
where $c_g(t), c_e(t)$ are the probability amplitudes of the initial (ground) state $|g\rangle$ and the target (excited) state $|e\rangle$, respectively. In this equation, $q$ is the constant Fano parameter and $D$ is the two-photon detuning,
\begin{equation}
\label{ionization_widths}
\Gamma_g=\Gamma^p_g+\Gamma^s_g, \quad \Gamma_e=\Gamma^p_e+\Gamma^s_e
\end{equation}
are the total ionization widths and
\begin{equation}
\label{ionization_widths}
\Sigma_g=\Sigma^p_g+\Sigma^s_g, \quad \Sigma_e=\Sigma^p_e+\Sigma^s_e
\end{equation}
are the corresponding dynamic Stark shifts of states $|g\rangle$ and $|e\rangle$, respectively. Note that the individual ionization widths and Stark shifts are proportional to the intensities of the pump and Stokes pulses
\begin{equation}
\Gamma^{\beta}_{\alpha}(t)=G^{\beta}_{\alpha}I_\beta(t), \quad \Sigma^{\beta}_{\alpha}(t)=S^{\beta}_{\alpha}I_\beta(t) \; (\alpha=g,e; \beta=p,s),
\end{equation}
where the coefficients $G^{\beta}_{\alpha}, S^{\beta}_{\alpha}$ depend on the particular atomic states and the laser frequencies.

Using the modified probability amplitudes defined by the following population preserving phase transformation \cite{Vitanov97}
\begin{equation}
b_{\alpha}(t)=c_{\alpha}(t)\exp\left\{i\int_{-\infty}^t[\Sigma_g(t')+\frac{1}{2}q\Gamma^p_g(t')]dt'\right\},
\end{equation}
where $\alpha=g,e$, we end up with the equation
\begin{eqnarray}
\label{system}
i
\left[
\begin{array}{c}
\dot{b}_g\\
\\
\dot{b}_e
\end{array}
\right]=&
\left[
\begin{array}{cc}
-\frac{1}{2}\Gamma^p_g(q+i)-\frac{1}{2}i\Gamma^s_g & -\frac{1}{2}\sqrt{\Gamma^p_g\Gamma^s_e}(q+i)\\
& \\
-\frac{1}{2}\sqrt{\Gamma^p_g\Gamma^s_e}(q+i) & -\frac{1}{2}\Gamma^s_e(q+i)-\frac{1}{2}i\Gamma^p_e+\delta
\end{array}
\right] \nonumber\\
\times\left[
\begin{array}{c}
b_g\\
\\
b_e
\end{array}
\right],&
\end{eqnarray}
where
\begin{equation}
\label{delta}
\delta=D+\Sigma_e-\Sigma_g-\frac{1}{2}q(\Gamma^p_g-\Gamma^s_e)
\end{equation}
is the effective two-photon detuning, which is in general time-dependent. Note that the validity of the two-level approximation is confirmed in Refs. \cite{Kylstra98,Paspalakis00}.

\begin{table}[t]
\caption{\label{tab:eff} Maximum excited-state population obtained with Gaussian STIRAP pulses (second column), the simple sin-cos protocol (third column), and optimal pulses (fourth column), for different values of parameter R (first column), expressing the strength of incoherent ionization. The top part of the table corresponds to zero effective detuning $\delta=0$, while the bottom part to nonzero $\delta\neq0$, given from Eq. (\ref{delta}) with $D=0$.}
\begin{ruledtabular}
\begin{tabular}{cccc}
\textrm{$R$}&
\textrm{Gaussian}&
\textrm{Sin-Cos}&
\textrm{Optimal}\\
\colrule
0 & 1 & 1 & 1 \\
\colrule
1/16 & 0.84 & 0.8332 & 0.8702 \\
\colrule
1/4 & 0.71 & 0.6945 & 0.7528 \\
\colrule
1 & 0.53 & 0.4347 & 0.5691 \\
\end{tabular}
\end{ruledtabular}
\begin{ruledtabular}
\begin{tabular}{cccc}
\textrm{$R$}&
\textrm{Gaussian}&
\textrm{Sin-Cos}&
\textrm{Optimal}\\
\colrule
0 & 0.53 & 0.5134 & 0.6280 \\
\colrule
1/16 & 0.51 & 0.4975 & 0.5948 \\
\colrule
1/4 & 0.48 & 0.4545 & 0.5373 \\
\colrule
1 & 0.40 & 0.3228 & 0.4207 \\
\end{tabular}
\end{ruledtabular}
\end{table}

As explained in detail in Ref. \cite{Vitanov97}, the terms $\Gamma^p_g$ (pump pulse applied on the $|g\rangle$-continuum transition) and $\Gamma^s_e$ (Stokes pulse applied on the $|e\rangle$-continuum transition) lead to the formation of a STIRAP system, where population is transferred from the ground to the excited state through the continuum. On the other hand, the terms $\Gamma^p_e$ (pump pulse applied on the $|e\rangle$-continuum transition) and $\Gamma^s_g$ (pump pulse applied on the $|g\rangle$-continuum transition) lead to irreversible ionization. At least one of these incoherent channels is always present, resulting to incomplete transfer of population between the bound states.

In Ref. \cite{Vitanov97} the authors use Gaussian pump and Stokes pulses of the same width $2T$ separated by a delay $2\tau$
\begin{equation}
\label{gaussian}
f_p(t)=e^{-\left(\frac{t-\tau}{T}\right)^2},\quad f_s(t)=e^{-\left(\frac{t+\tau}{T}\right)^2},
\end{equation}
and test the performance using the following ionization widths and Stark shifts
\begin{subequations}
\label{widths_shifts}
\begin{eqnarray}
\Gamma^p_g(t)=Af_p(t),&\quad &\Gamma^s_g(t)=0,\\
\Gamma^p_e(t)=RAf_p(t),&\quad &\Gamma^s_e(t)=Af_s(t),\\
\Sigma^p_g(t)=Af_p(t),&\quad &\Sigma^s_g(t)=-Af_s(t),\\
\Sigma^p_e(t)=Af_p(t),&\quad& \Sigma^s_e(t)=3Af_s(t),
\end{eqnarray}
\end{subequations}
while the Fano parameter is set to $q=-6$. Note that this is close to the value $q=-5.87$, corresponding to the hydrogen atom \cite{Yatsenko97}. Parameter $A$ is proportional to the intensity of the lasers, while $R$ quantifies the strength of incoherent ionization. Four values of parameter $R$ are used ($R=0, 1/16, 1/4, 1$), covering the range from weak to strong incoherent ionization, for both zero ($\delta=0$) and nonzero ($\delta\neq0$) effective two-photon detuning, given from Eq. (\ref{delta}) with $D=0$. Note that effective two-photon resonance can be achieved with additional laser pulses \cite{Yatsenko97} or frequency chirping \cite{Paspalakis97}. For each of these $4\times 2=8$ cases, the authors of Ref. \cite{Vitanov97} simulate Eq. (\ref{system}) for various widths and delays of the Gaussian pulses, and find the maximum excited-state population achieved. These results are summarized in the second column of Table \ref{tab:eff}.

\section{A simple sin-cos control protocol}

\label{sec:linear}

Let us for a moment consider the idealized situation without incoherent ionization and with zero effective two-photon detuning, $R=0$ and $\delta=0$, and set
\begin{equation}
\label{Gamma_sincos}
\Gamma^p_g(t)=A\sin^2{\theta(t)},\quad \Gamma^s_e(t)=A\cos^2{\theta(t)},
\end{equation}
so the mixing angle $\theta(t)$ is defined as $\tan{\theta(t)}=\sqrt{\Gamma^p_g(t)/\Gamma^s_e(t)}$. In this case, the adiabatic eigenstates of Eq. (\ref{system}) are
\begin{equation}
\label{eigenstates}
|\psi_0\rangle=\left[
\begin{array}{c}
  \cos{\theta}\\
  -\sin{\theta}
\end{array}
\right],
\quad
|\psi_1\rangle=\left[
\begin{array}{c}
  \sin{\theta}\\
  \cos{\theta}
\end{array}
\right],
\end{equation}
with corresponding eigenvalues $\omega_0=0$ and $\omega_1/A=-(q+i)/2$. It can be easily shown that the probability amplitudes in the adiabatic basis,
\begin{equation}
\label{transformation}
\left[
\begin{array}{c}
  a_0\\
  a_1
\end{array}
\right]
=
\left[
\begin{array}{cc}
  \cos{\theta} & -\sin{\theta}\\
  \sin{\theta} & \cos{\theta}
\end{array}
\right]
\left[
\begin{array}{c}
  b_g\\
  b_e
\end{array}
\right],
\end{equation}
obey the following equation
\begin{equation}
\label{adiabatic}
i
\left[
\begin{array}{c}
  \dot{a}_0\\
  \dot{a}_1
\end{array}
\right]
=
\left[
\begin{array}{cc}
  0 & -i\dot{\theta}\\
  i\dot{\theta} & -\frac{A}{2}(q+i)
\end{array}
\right]
\left[
\begin{array}{c}
  \dot{a}_0\\
  \dot{a}_1
\end{array}
\right].
\end{equation}
Observe that, if the mixing angle is slowly varied from $\theta(0)=0$ to $\theta(T)=\pi/2$ at the final time $t=T$, then the system remains in the eigenstate $|\psi_0\rangle$, which is gradually transformed from $|\psi_0(0)\rangle=|g\rangle$ to $|\psi_0(T)\rangle=|e\rangle$, thus a perfect population transfer is accomplished. This is the reason behind the success of the Gaussian STIRAP pulses used in Ref. \cite{Vitanov97}. The adiabaticity condition is satisfied when $\dot{\theta}/A\ll q/2$, where observe from Eq. (\ref{adiabatic}) that $A|q|/2$ is the frequency separation between the adiabatic eigenstates, thus larger Fano factors facilitate the adiabatic process. Of course, deviations from the adiabaticity and/or the ideal conditions $R=0, \delta=0$ lead to incomplete population transfer, see the second column of Table \ref{tab:eff}.

\begin{figure}[t]
 \centering
		\begin{tabular}{c}
      \subfigure[\quad\quad\quad\quad\quad\quad\quad\quad\quad\quad\quad\quad\quad\quad\quad\quad]{
	            \label{fig:Ramp_T}
	            \includegraphics[width=.85\linewidth]{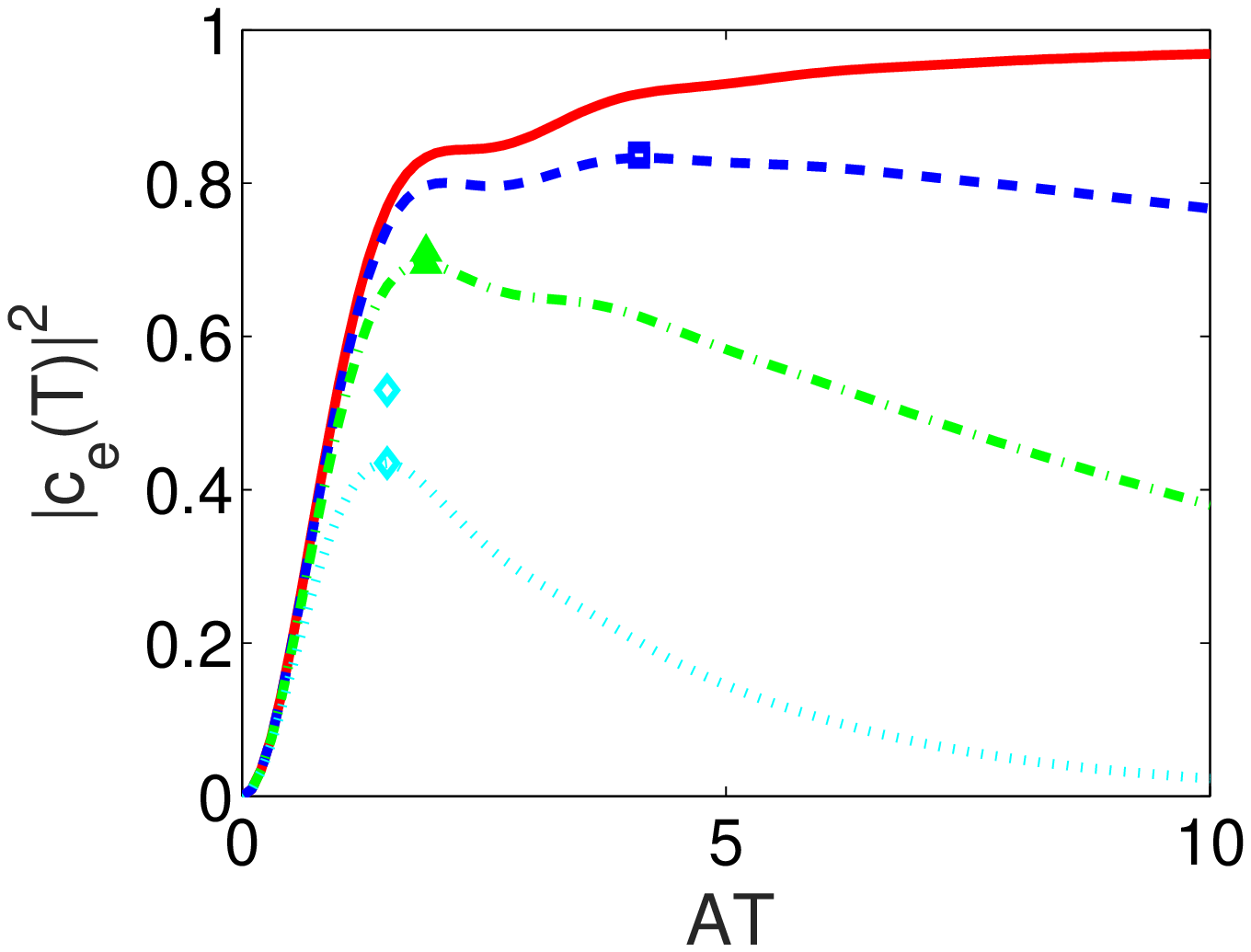}} \\
      \subfigure[\quad\quad\quad\quad\quad\quad\quad\quad\quad\quad\quad\quad\quad\quad\quad\quad]{
	            \label{fig:Ramp_T_full}
	            \includegraphics[width=.85\linewidth]{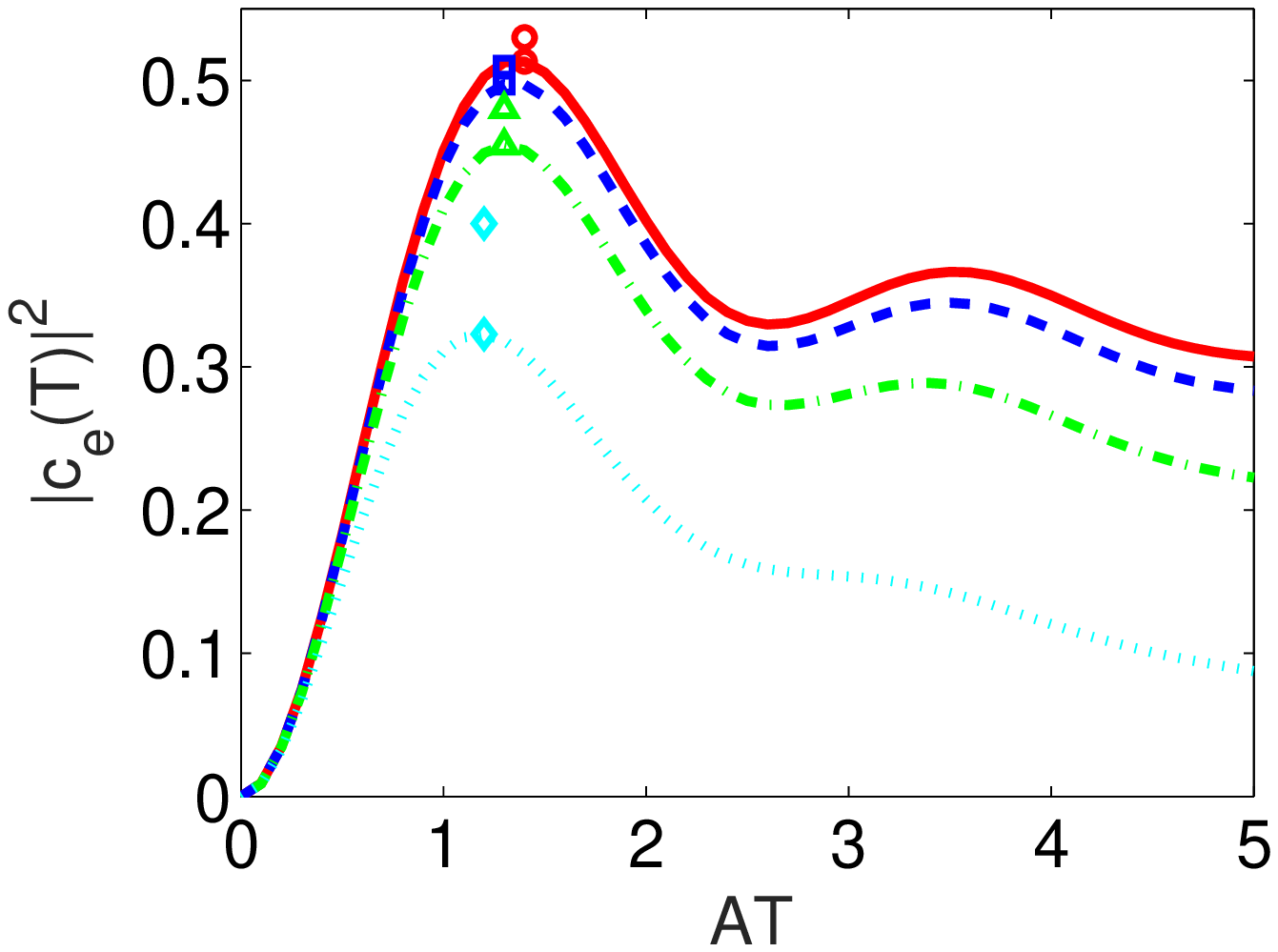}}
		\end{tabular}
\caption{(Color online) Excited-state population obtained with the simple sin-cos protocol as a function of the normalized pulse duration $AT$, for $q=-6$ and four different values of parameter $R$, which essentially determines the strength of incoherent ionization, $R=0$ (red solid line), $R=1/16$ (blue dashed line), $R=1/4$ (green dashed-dotted line), $R=1$ (cyan dotted line). (a) Case with effective two-photon resonance $\delta=0$. (b) Case with nonzero $\delta$ obtained from Eq. (\ref{delta}). On each curve is highlighted the maximum efficiency achieved, except case $\delta=0, R=0$, where complete population transfer is obtained in the limit of large $AT$. The isolated points with the same abscissas indicate the best efficiencies obtained in Ref. \cite{Vitanov97} with Gaussian STIRAP pulses for the same values of $R$.}
\label{fig:Eff_vs_ramp}
\end{figure}

The form of Eq. (\ref{adiabatic}) motivates us to consider the situation where the mixing angle increases with constant rate \cite{Carroll92}
\begin{equation}
\dot{\theta}=\frac{\pi}{2T} \quad \mbox{constant},
\end{equation}
thus the ionization widths of Eq. (\ref{Gamma_sincos}) follow a simple sin-cos control protocol
\begin{equation}
\label{sincos}
\Gamma^p_g(t)=A\sin^2{\left(\frac{\pi t}{2T}\right)},\quad \Gamma^s_e(t)=A\cos^2{\left(\frac{\pi t}{2T}\right)}.
\end{equation}
The usage of this protocol is also motivated by our recent work \cite{Stefanatos_Opt_Lett20} in a different context, involving the double-$\Lambda$ atom–light coupling scheme. We show there that the performance of the aforementioned protocol, when applied to the dynamical system (\ref{system}) with $R=q=\delta=0$, approaches that of the optimal protocol, where the mixing angle is varied linearly too, but also includes some initial and final $\delta$-kicks changing $\theta$ instantaneously at the beginning and end.
For the sin-cos protocol, Eq. (\ref{adiabatic}) can be easily integrated and we obtain in the idealized situation the following analytic expression for the population of the excited state at the final time $t=T$
\begin{equation}
\label{constant_eff}
|c_e(T)|^2=e^{-\eta AT}\left[\cosh{(\kappa AT)}+\frac{\eta}{2\kappa}\sinh{(\kappa AT)}\right]^2,
\end{equation}
where
\begin{equation}
\label{eta_kappa}
\eta=\frac{1}{2}(1-iq),\quad\kappa=\frac{1}{2}\sqrt{\eta^2-\left(\frac{\pi}{AT}\right)^2}.
\end{equation}

\begin{figure*}[t]
 \centering
		\begin{tabular}{cc}    	
      \subfigure[\quad\quad\quad\quad\quad\quad\quad\quad\quad\quad\quad\quad\quad\quad\quad\quad]{
	            \label{fig:Con_lin_d0}
	            \includegraphics[width=.4\linewidth]{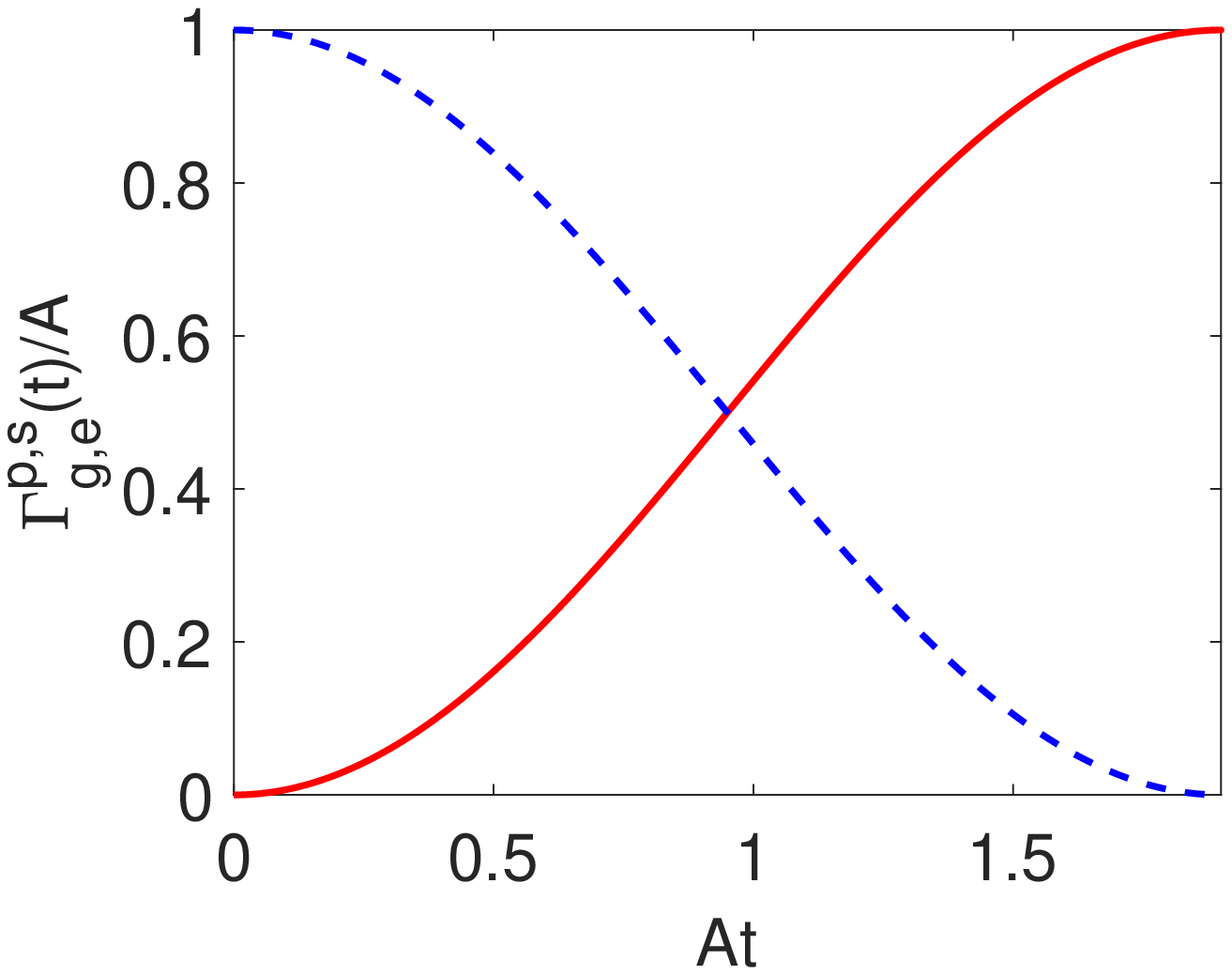}} &
      \subfigure[\quad\quad\quad\quad\quad\quad\quad\quad\quad\quad\quad\quad\quad\quad\quad\quad]{
	            \label{fig:Pop_lin_d0}
	            \includegraphics[width=.4\linewidth]{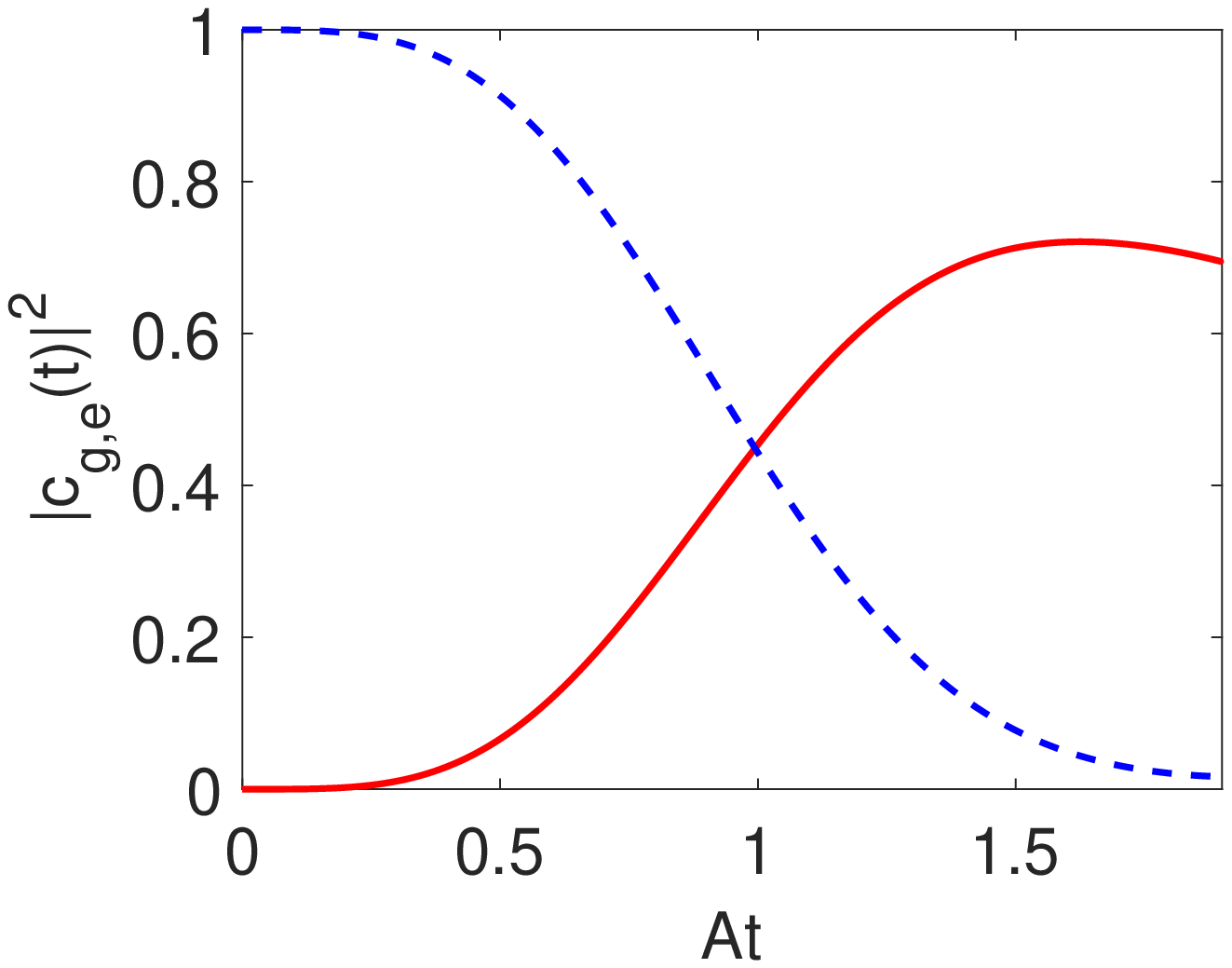}} \\
       \subfigure[\quad\quad\quad\quad\quad\quad\quad\quad\quad\quad\quad\quad\quad\quad\quad\quad]{
	            \label{fig:Con_lin_d}
	            \includegraphics[width=.4\linewidth]{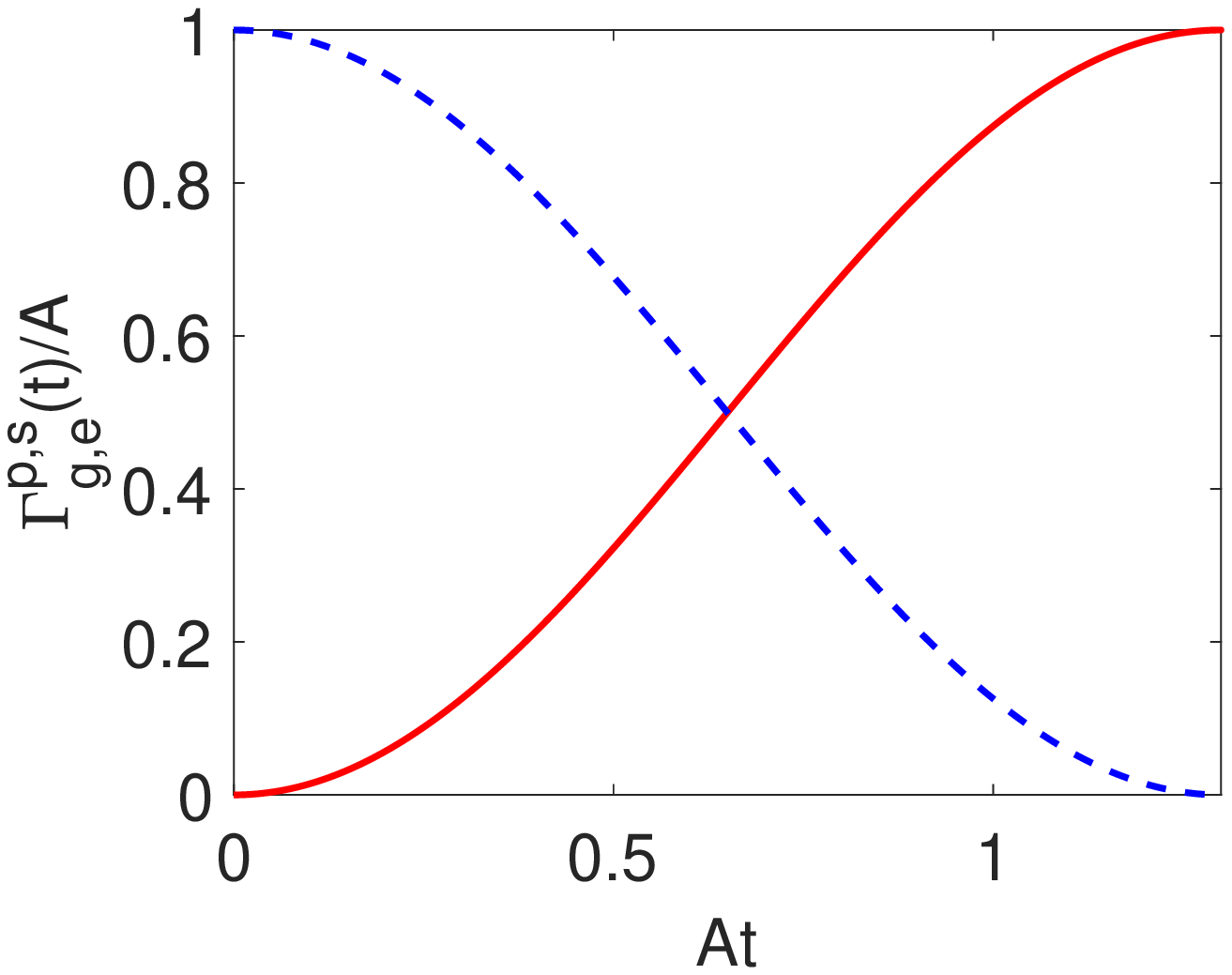}} &
      \subfigure[\quad\quad\quad\quad\quad\quad\quad\quad\quad\quad\quad\quad\quad\quad\quad\quad]{
	            \label{fig:Pop_lin_d}
	            \includegraphics[width=.4\linewidth]{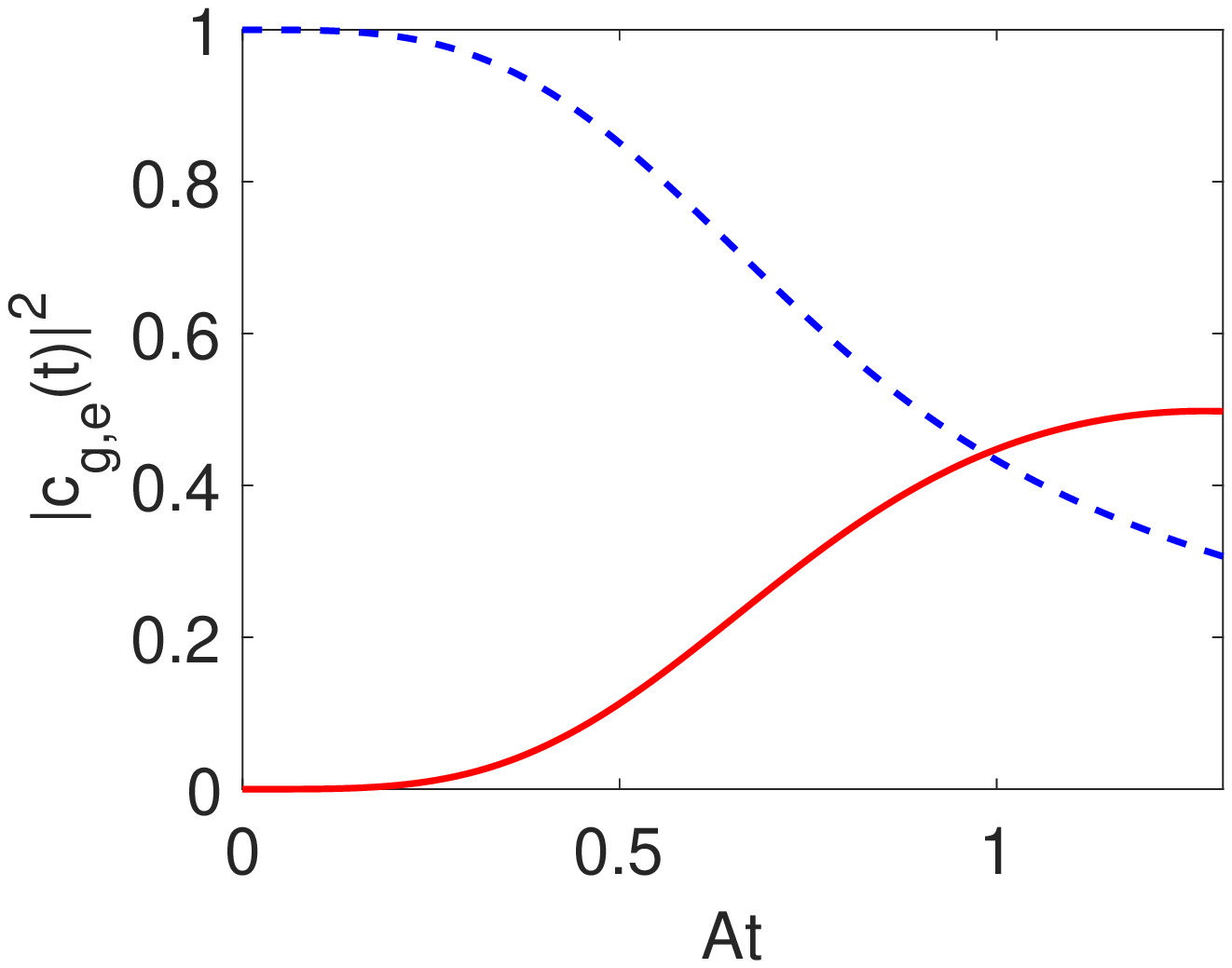}}
		\end{tabular}
\caption{(Color online) (a) Ionization pulses $\Gamma^p_g(t)$ (red solid line) and $\Gamma^s_e(t)$ (blue dashed line) obtained with the simple sin-cos protocol for normalized duration $AT=1.9$, in the case of effective two-photon resonance $\delta=0$ and $R=1/4$. (b) Corresponding evolution of populations of the ground (blue dashed line) and excited states (red solid line). (c) Ionization pulses obtained with the sin-cos protocol for normalized duration $AT=1.3$, $\delta\neq 0$ and $R=1/16$. (d) Corresponding evolution of populations.}
\label{fig:Ex1}
\end{figure*}

The transfer efficiency of the sin-cos protocol for the idealized case $R=0, \delta=0$, given in Eq. (\ref{constant_eff}), is displayed in Fig. \ref{fig:Ramp_T} (red solid line) as a function of the normalized duration $AT$, a quantity which is also proportional to the area ($AT/2$) of the pulses (\ref{sincos}). Observe that the efficiency increases with the pulse area and approaches unity in the limit $AT\rightarrow\infty$. We also display the efficiency for the nonzero values of $R$ previously used, and observe that in this case the efficiency is maximized at a finite pulse area. Both characteristics are also observed with Gaussian pulses in Ref. \cite{Vitanov97}. The maximum efficiencies corresponding to different $R$ are shown with markers in Fig. \ref{fig:Ramp_T}, and are also listed in the third column of the upper part of Table \ref{tab:eff}. For comparison, we also display with markers the maximum efficiencies obtained for the same values of $R$ in Ref. \cite{Vitanov97} using Gaussian pulses, listed in the second column of the upper part of Table \ref{tab:eff}. From Fig. \ref{fig:Ramp_T} or Table \ref{tab:eff} we note that the two methods achieve the same maximum efficiency for $R=0$, similar maximum efficiencies for $R=1/16$ and $R=1/4$ (blue squares and green triangles respectively), while for $R=1$ (cyan diamonds) the maximum obtained with Gaussian pulses is larger. In Fig. \ref{fig:Ramp_T_full} we display similar results but for the case where $\delta\neq0$, obtained from Eq. (\ref{delta}) with $D=0$. In this case, the maximum efficiencies obtained with the sin-cos protocol are smaller than those obtained with Gaussian pulses for all values of $R$, with the difference becoming more distinct for larger $R$. The maximum efficiencies obtained with the two protocols for different $R$ are also listed in the second and third columns of the lower part of Table \ref{tab:eff}. The better performance of the Gaussian pulses can be attributed to the fact that in this case there are two parameters over which the optimization is performed, the pulse width $2T$ and the delay $2\tau$, while for the sin-cos pulses the only parameter is the pulse duration $T$. In Fig. \ref{fig:Ex1} we show the sin-cos ionization widths (\ref{sincos}) and the corresponding evolution of populations for two specific examples, $\delta=0, R=1/4, AT=1.9$ (first row) and $\delta\neq 0, R=1/16, AT=1.3$ (second row).

\section{Optimal control analysis of the problem}

\label{optimal_solution}

In order to find pulse shapes which outperform the Gaussian pulses, we employ in this section optimal control theory \cite{Bryson}. We use as state variables $x_i$, $i=1,2,3,4$, the real and imaginary parts of the probability amplitudes $b_g, b_e$,
\begin{equation}
\label{state}
b_g=x_1+ix_2,\quad b_e=x_3+ix_4
\end{equation}
and as control variables the square roots of the ionization widths,
\begin{equation}
\label{controls}
\sqrt{\Gamma^p_g(t)}=u_1(t),\quad \sqrt{\Gamma^s_e(t)}=u_2(t).
\end{equation}
With the above definitions, we find from Eq. (\ref{system}), using also Eq. (\ref{widths_shifts}), the state equations
\begin{subequations}
\label{state_system}
\begin{eqnarray}
\dot{x}_1&=&-\frac{1}{2}u_1^2x_1-\frac{q}{2}u_1^2x_2-\frac{1}{2}u_1u_2x_3-\frac{q}{2}u_1u_2x_4, \\
\dot{x}_2&=&\frac{q}{2}u_1^2x_1-\frac{1}{2}u_1^2x_2+\frac{q}{2}u_1u_2x_3-\frac{1}{2}u_1u_2x_4,  \\
\dot{x}_3&=&-\frac{1}{2}u_1u_2x_1-\frac{q}{2}u_1u_2x_2-\frac{1}{2}(Ru_1^2+u_2^2)x_3\nonumber\\
&&+\left(\delta-\frac{q}{2}u_2^2\right)x_4, \\
\dot{x}_4&=&\frac{q}{2}u_1u_2x_1-\frac{1}{2}u_1u_2x_2+\left(\frac{q}{2}u_2^2-\delta\right)x_3\nonumber\\
&&-\frac{1}{2}(Ru_1^2+u_2^2)x_4,
\end{eqnarray}
\end{subequations}
where for the effective two-photon detuning we distinguish two cases as before, one with $\delta=0$ and one with $\delta\neq 0$ obtained from Eq. (\ref{delta}) for $D=0$ which, using Eqs. (\ref{widths_shifts}) and (\ref{controls}), becomes
\begin{equation}
\label{d}
\delta=-\frac{q}{2}u_1^2+\left(4+\frac{q}{2}\right)u_2^2.
\end{equation}
We would like to find the controls $u_1(t), u_2(t)$ which maximize the final population of the excited state
\begin{equation}
\label{final_exc}
|c_e(T)|^2=|b_e(T)|^2=x_3(T)^2+x_4(T)^2
\end{equation}
when starting from the ground state $b_g(0)=1$, corresponding to the initial conditions
\begin{equation}
\label{initial_condition}
x_1(0)=1,\quad x_2(0)=x_3(0)=x_4(0)=0,
\end{equation}
while satisfying the constraints
\begin{equation}
\label{constraint}
0\leq u_i(t)/\sqrt{A} \leq 1,\quad i=1,2.
\end{equation}

An important observation can be immediately made by inspecting the system equations. If $\pi_1, \pi_2$ denote the optimal policies in the time intervals $[0,\,T_1]$ and $[0,\,T_2]$, respectively, with $T_1\leq T_2$, then $|c_e(T_2)|^2\geq |c_e(T_1)|^2$, i.e. a better transfer efficiency can be obtained in a longer duration. This can be easily shown as follows. Consider the longer time interval and suppose that for $0\leq t\leq T_1$ we apply policy $\pi_1$, while for $T_1< t\leq T_2$ we set $u_1(t)=u_2(t)=0$. But with this latter control choice, the left hand sides of system equations (\ref{state_system}) become zero, for both $\delta=0$ and $\delta$ given from Eq. (\ref{d}), thus the system remains in the state reached at $t=T_1$ and the efficiency at the final time $t=T_2$ equals $|c_e(T_1)|^2$, the efficiency of policy $\pi_1$. Obviously, this should be lower or equal to the efficiency obtained with policy $\pi_2$, which is by definition optimal over the whole time interval $[0,\,T_2]$. Note that there is no contradiction with the sin-cos protocol and the Gaussian pulses, where the maximum efficiency is obtained for finite duration (pulse area), since for the optimal pulses there is no association between the pulse duration and area.

In order to obtain an idea about the form of the optimal $u_i(t)$, we will use some simple elements from optimal control theory. The control Hamiltonian of the problem is defined as \cite{Bryson}
\begin{equation}
\mathcal{H}_c=\lambda_1\dot{x}_1+\lambda_2\dot{x}_2+\lambda_3\dot{x}_3+\lambda_4\dot{x}_4=\mathcal{H}_c(\bm{\lambda},\mathbf{x},\mathbf{u}),
\end{equation}
which becomes a function of the state variables $\mathbf{x}=[x_1, x_2, x_3, x_4]^T$  and the controls $\mathbf{u}=[u_1, u_2]^T$ by replacing the state derivatives in the above definition using Eq. (\ref{state_system}).
The Lagrange multipliers $\bm{\lambda}=[\lambda_1, \lambda_2, \lambda_3, \lambda_4]^T$ satisfy the adjoint equations
\begin{equation}
\label{adjoint}
\dot{\bm{\lambda}} = -\frac{\partial \mathcal{H}_c}{\partial\mathbf{x}}.
\end{equation}
and the terminal conditions \cite{Bryson}
\begin{subequations}
\label{final_lambda}
\begin{eqnarray}
\lambda_1(T)&=&\frac{\partial |c_e(T)|^2}{\partial x_1(T)}=0,\\
\lambda_2(T)&=&\frac{\partial |c_e(T)|^2}{\partial x_2(T)}=0,\\
\lambda_3(T)&=&\frac{\partial |c_e(T)|^2}{\partial x_3(T)}=2x_3(T),\\
\lambda_4(T)&=&\frac{\partial |c_e(T)|^2}{\partial x_4(T)}=2x_4(T).
\end{eqnarray}
\end{subequations}

According to Pontryagin's Maximum Principle \cite{Bryson}, the optimal controls are chosen to maximize $\mathcal{H}_c$. But from Eq. (\ref{system}) it turns out that $\mathcal{H}_c$ is a quadratic function of the control variables $u_1, u_2$, which are restricted in the square (\ref{constraint}) on the $u_1u_2$-plane. If for some finite time interval the optimal $u_i$ is one of the bounds of the constraint (\ref{constraint}) then it is called a bang control, otherwise it lies in the interior and is determined from the relation $\partial \mathcal{H}_c/\partial u_i=0$. From our previous experience with systems where $\mathcal{H}_c$ is quadratic in the controls \cite{Stefanatos_PRA04}, in the context of nuclear magnetic resonance spectroscopy, we know that for short durations $T$ the optimal $u_i$ have the bang form (both obtain the maximum value), since in this case the major limitation is the short available time and not the incoherent losses. Of course, the transfer efficiencies obtained are also limited. For the more interesting case with longer durations, where larger efficiencies can be achieved, the optimal $u_i$ assume the bang-interior and interior-bang form, in order to engineer a path along which the ionization losses are minimized (note that the bang segments correspond to maxima). The analytical determination of the switching times, from bang to interior and vice versa, is a formidable task, as is the solution of the optimal control problem which is a two-point boundary value problem, since we are given the initial conditions (\ref{initial_condition}) for the state variables and the terminal conditions (\ref{final_lambda}) for the Lagrange multipliers. For example, we mention that even in Ref. \cite{Stefanatos_PRA04}, where we indeed solve a (simpler) problem of this type, the switching times are determined by solving a system of transcendental equations, which eventually needs to be done numerically. For these reasons, we will not pursue further the analytical investigation of the problem, but continue using numerical optimal control in the next section.

\section{Numerical results and discussion}

\label{sec:results}

\begin{figure}[t]
 \centering
		\begin{tabular}{c}
     	
      \subfigure[\quad\quad\quad\quad\quad\quad\quad\quad\quad\quad\quad\quad\quad\quad\quad\quad]{
	            \label{fig:Eff_T}
	            \includegraphics[width=.85\linewidth]{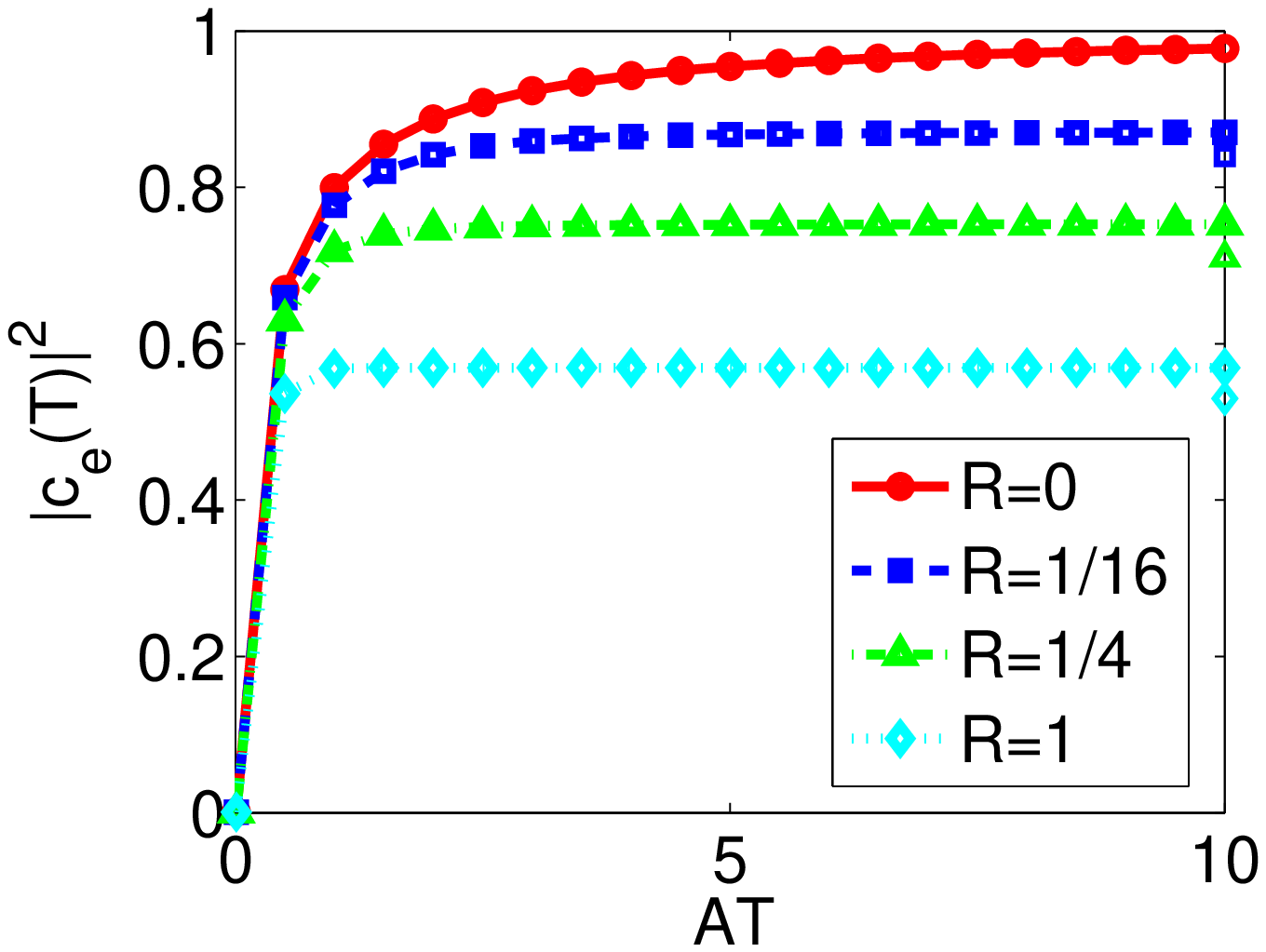}} \\
      \subfigure[\quad\quad\quad\quad\quad\quad\quad\quad\quad\quad\quad\quad\quad\quad\quad\quad]{
	            \label{fig:Eff_T_full}
	            \includegraphics[width=.85\linewidth]{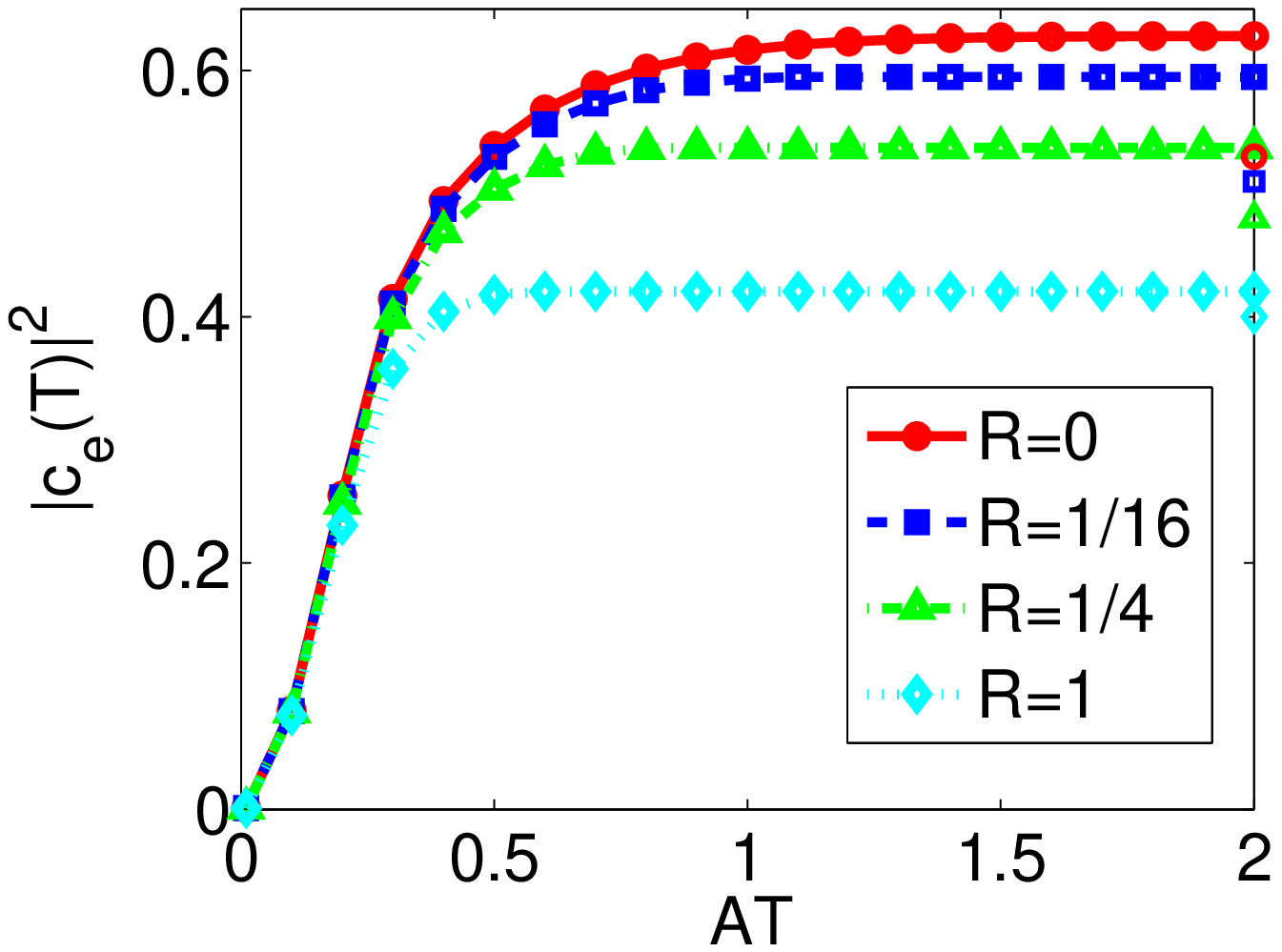}}
		\end{tabular}
\caption{(Color online) Excited-state population obtained with the optimal pulses as a function of the normalized pulse duration $AT$, with step $\delta T=0.1/A$, for $q=-6$ and four different values of parameter $R$, which essentially determines the strength of incoherent ionization. The isolated points on the right vertical axes indicate the best efficiencies obtained in Ref. \cite{Vitanov97} with Gaussian STIRAP pulses for the same values of $R$. (a) Case with effective two-photon resonance $\delta=0$. (b) Case with nonzero $\delta$ obtained from Eq. (\ref{d}).}
\label{fig:Eff_vs_T}
\end{figure}

In order to solve numerically the control problem defined in the previous section, we use the optimal control solver BOCOP \cite{bocop}. In Fig. \ref{fig:Eff_T} we plot for the case $\delta=0$ the final population of the excited state $|c_e(T)|^2$ obtained with optimal pulses, as a function of the normalized duration $AT$ with step $\delta T=0.1/A$, using $q=-6$ and the four values of parameter $R$ previously utilized. Observe that in all cases the efficiency increases with increasing duration, as it was proved in the previous section. For $R=0$ the efficiency approaches unity for large $T$, as is the case for the Gaussian pulses and the sin-cos protocol. For $R>0$, the efficiency saturates to a value lower than one, which decreases with increasing $R$. Note that efficiencies close to these saturation limits can be obtained at finite durations, which become smaller for increasing $R$. We list the saturation limits (maximum efficiencies) in the fourth column of the upper part of Table \ref{tab:eff}. For comparison, in Fig. \ref{fig:Eff_T} we display with isolated points on the right vertical axes the best efficiencies obtained in Ref. \cite{Vitanov97} with Gaussian pulses, for the same values of $R$. Although for $R=0$ both methods have the same maximum efficiency (unity), the optimal method performs better for increasing $R>0$. Similar results are obtained for the case $\delta\neq 0$, shown in Fig. \ref{fig:Eff_T_full}, but now the saturation efficiencies are smaller than before, for the same values of $R$, and are obtained in shorter durations. As before, these maximum efficiencies are listed in the fourth column of the lower part of Table \ref{tab:eff}. Observe that, even for $R=0$, the saturation efficiency is less than unity. The optimal method performs better for all $R$, but now the improvement is reduced for increasing $R$.

\begin{figure}[t]
\centering
\includegraphics[width=0.85\linewidth]{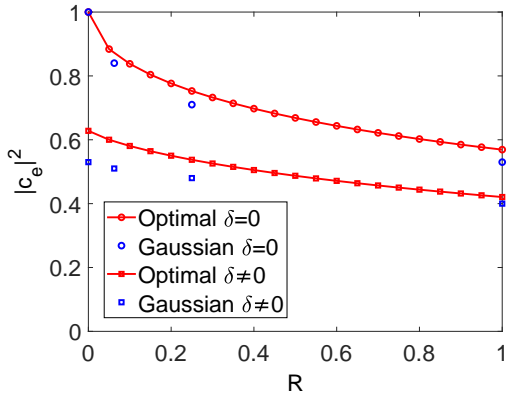}
\caption{(Color online) Maximum excited-state population obtained with the optimal pulses (in the limit of large $AT$) as a function of parameter $R$, with step $\delta R=0.05$, for the case of effective two-photon resonance $\delta=0$ (red circles) and nonzero $\delta$ from Eq. (\ref{d}) (red squares). The isolated points at the specific values $R=0, 1/16, 1/4, 1$ indicate the best efficiencies obtained in Ref. \cite{Vitanov97} using Gaussian STIRAP pulses.}
\label{fig:Eff_R}
\end{figure}

\begin{figure}[t]
 \centering
		\begin{tabular}{c}
     	
      \subfigure[\quad\quad\quad\quad\quad\quad\quad\quad\quad\quad\quad\quad\quad\quad\quad\quad]{
	            \label{fig:Eff_q}
	            \includegraphics[width=.85\linewidth]{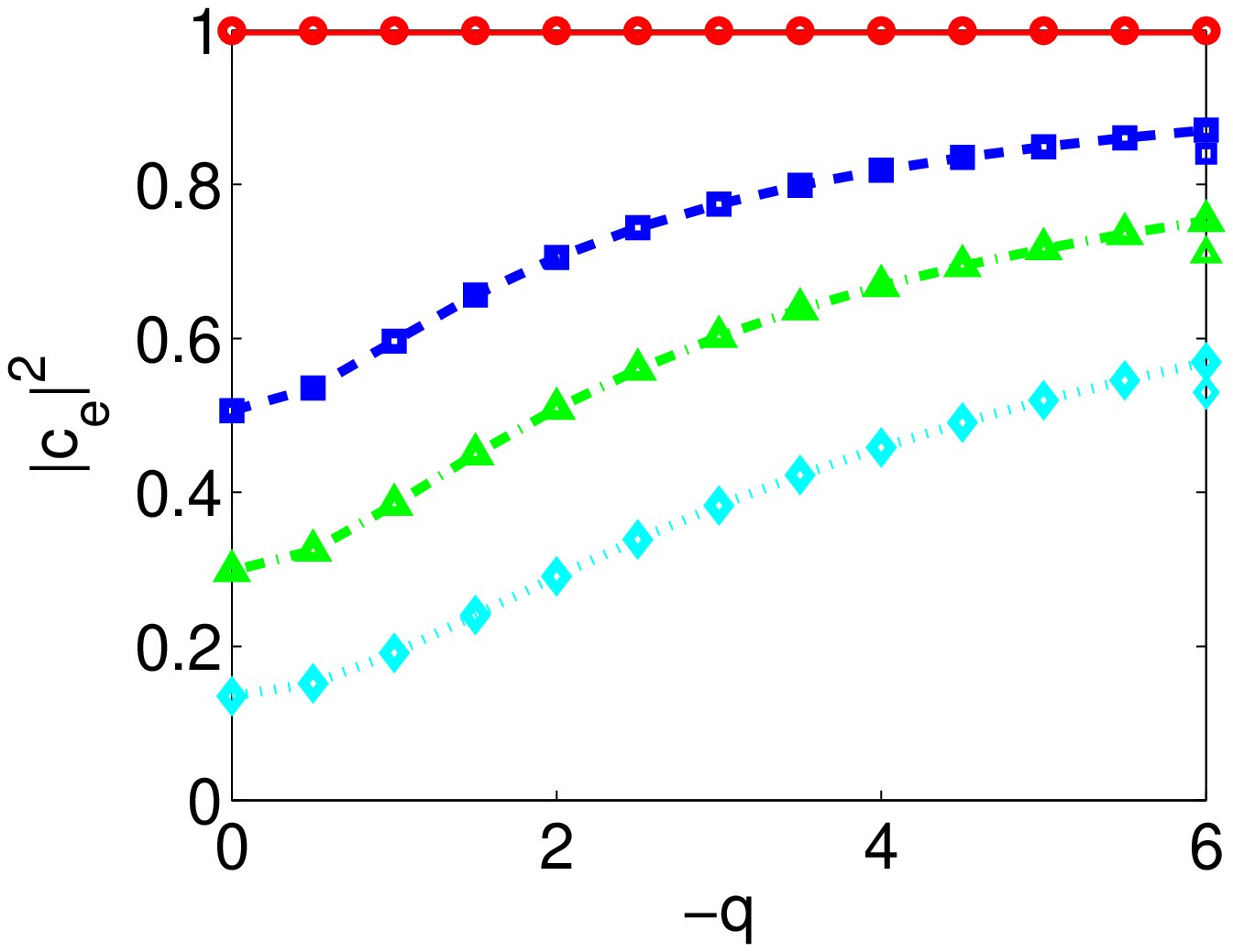}} \\
      \subfigure[\quad\quad\quad\quad\quad\quad\quad\quad\quad\quad\quad\quad\quad\quad\quad\quad]{
	            \label{fig:Eff_q_full}
	            \includegraphics[width=.85\linewidth]{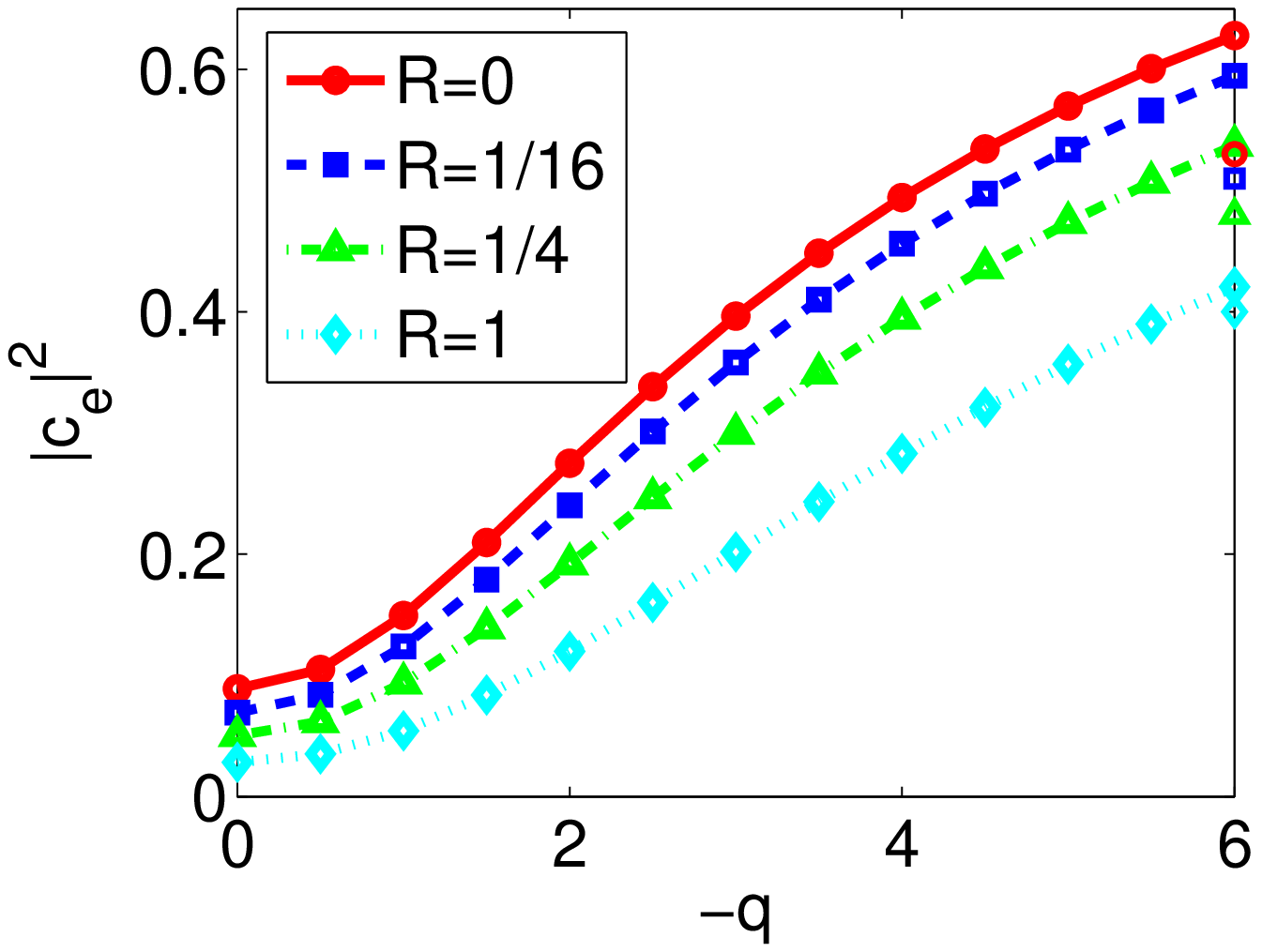}}
		\end{tabular}
\caption{(Color online) Maximum excited-state population obtained with the optimal pulses (in the limit of large $AT$) as a function of Fano parameter $q$, with step $\delta q =0.5$, for four different values of $R$. The isolated points at $q=-6$ indicate the best efficiencies obtained in Ref. \cite{Vitanov97} with Gaussian STIRAP pulses for the same values of $R$. (a) Case with effective two-photon resonance $\delta=0$. (b) Case with nonzero $\delta$ obtained from Eq. (\ref{d}).}
\label{fig:Eff_vs_q}
\end{figure}

\begin{figure*}[t]
 \centering
		\begin{tabular}{cc}
     	
      \subfigure[\quad\quad\quad\quad\quad\quad\quad\quad\quad\quad\quad\quad\quad\quad\quad\quad]{
	            \label{fig:Con_opt_d0}
	            \includegraphics[width=.4\linewidth]{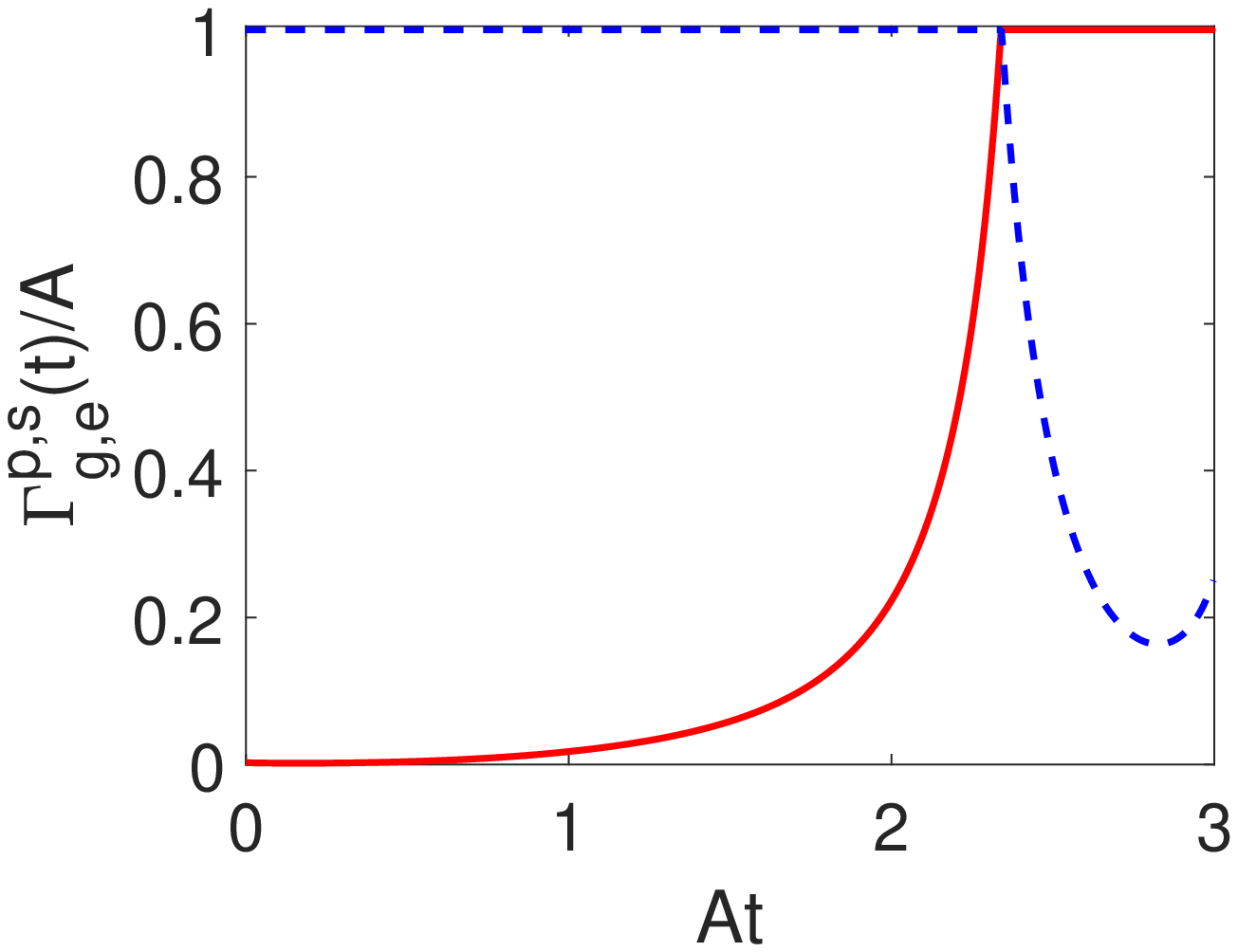}} &
      \subfigure[\quad\quad\quad\quad\quad\quad\quad\quad\quad\quad\quad\quad\quad\quad\quad\quad]{
	            \label{fig:Pop_opt_d0}
	            \includegraphics[width=.4\linewidth]{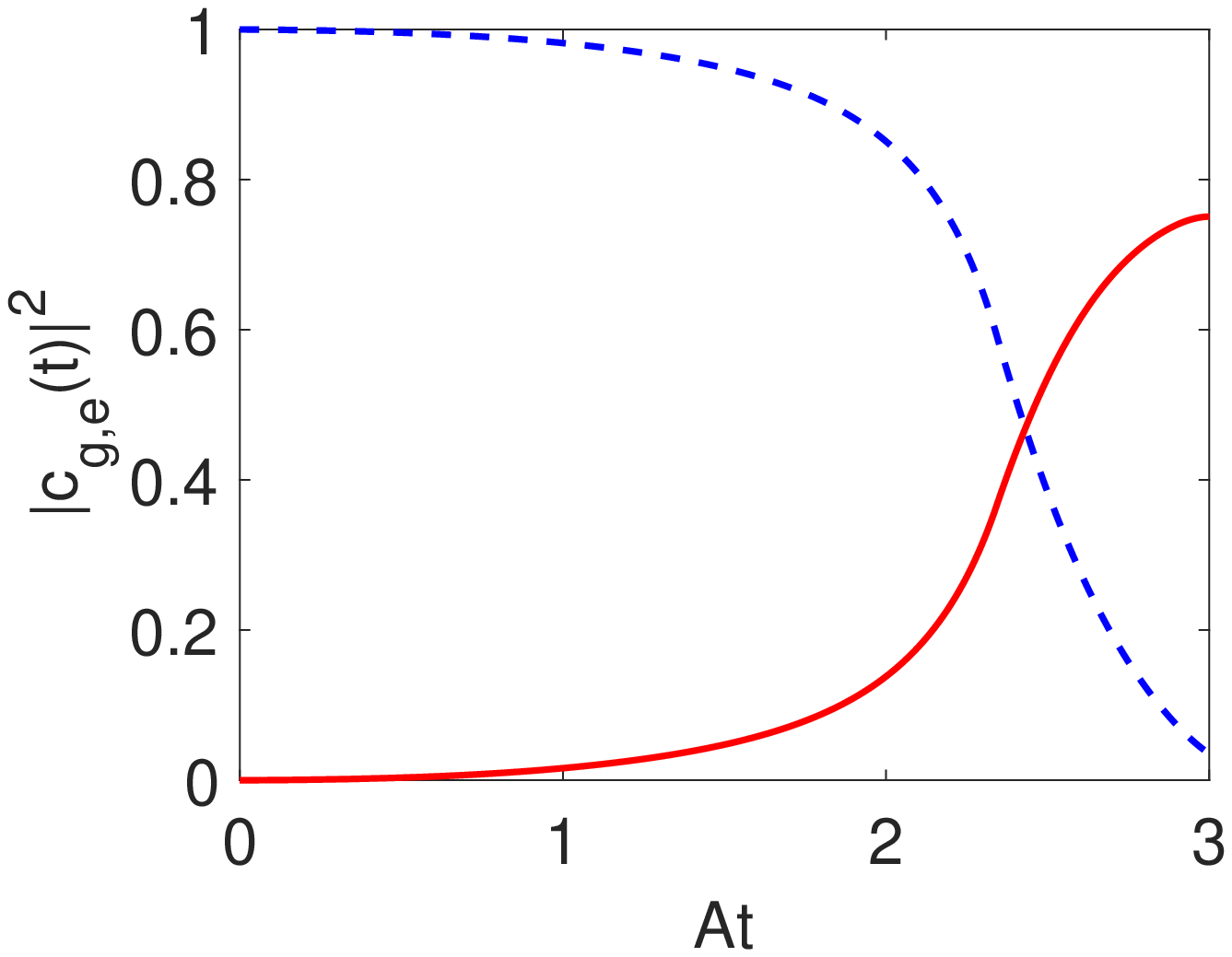}} \\
\subfigure[\quad\quad\quad\quad\quad\quad\quad\quad\quad\quad\quad\quad\quad\quad\quad\quad]{
	            \label{fig:Con_opt_d}
	            \includegraphics[width=.4\linewidth]{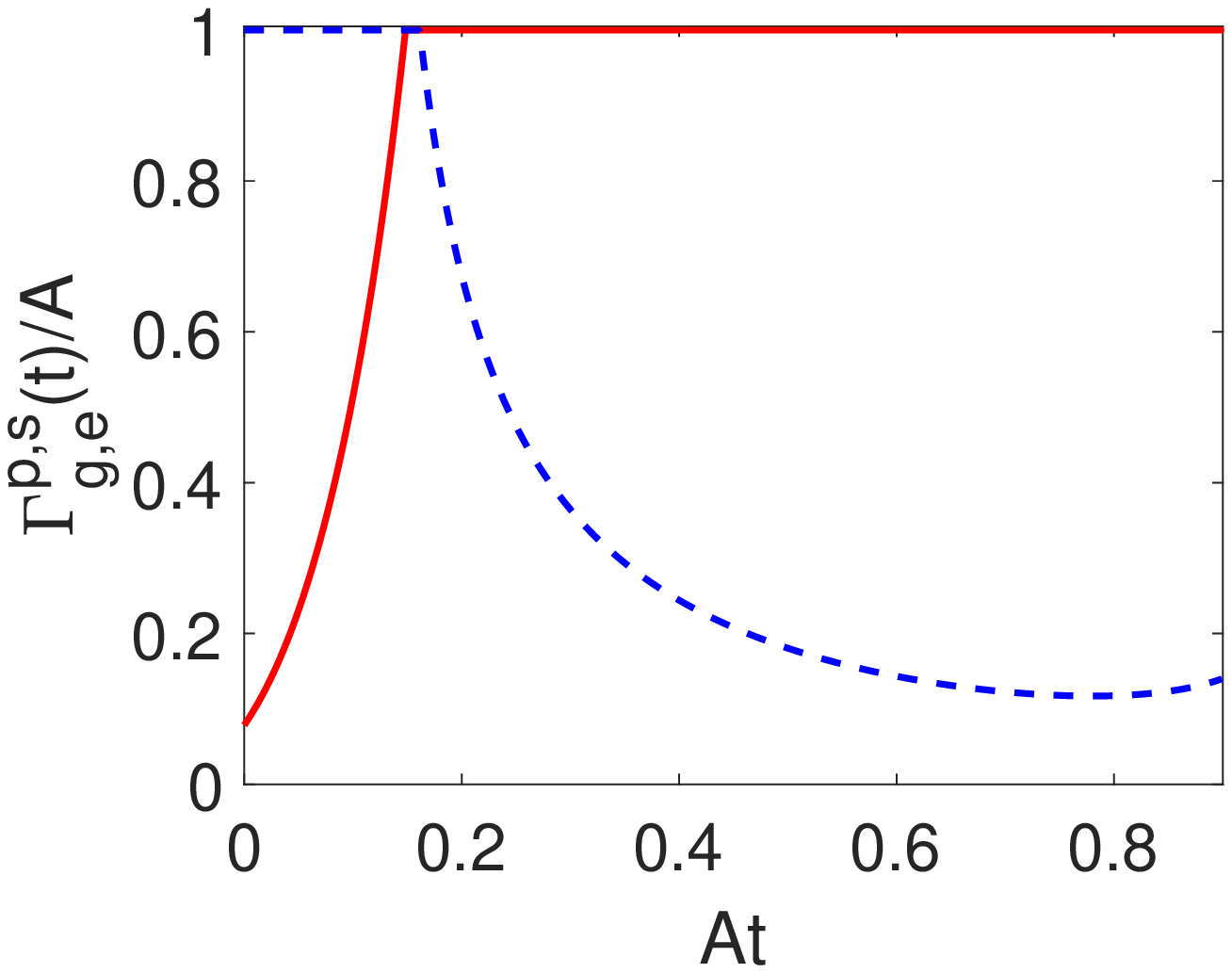}} &
      \subfigure[\quad\quad\quad\quad\quad\quad\quad\quad\quad\quad\quad\quad\quad\quad\quad\quad]{
	            \label{fig:Pop_opt_d}
	            \includegraphics[width=.4\linewidth]{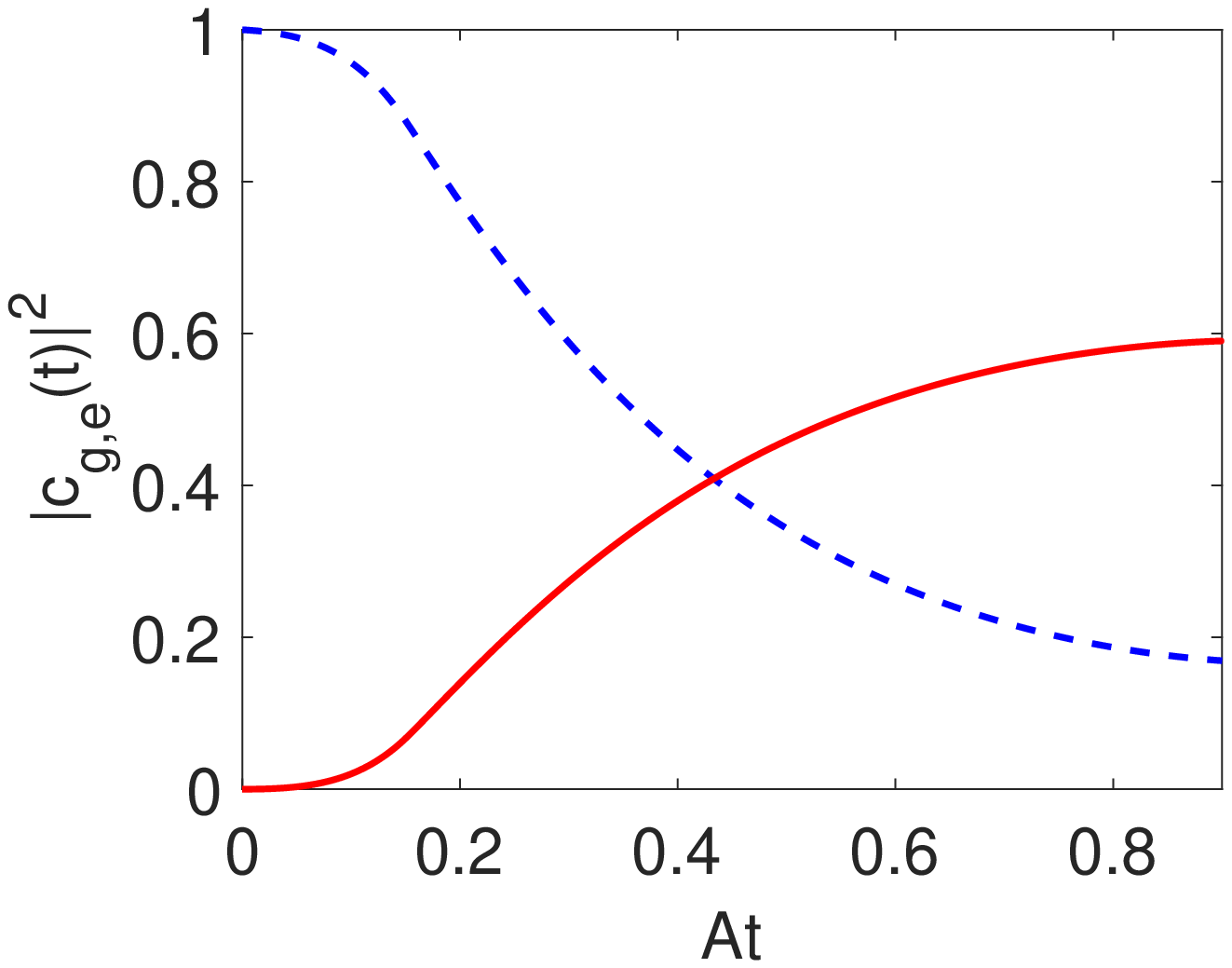}} \\
      \subfigure[\quad\quad\quad\quad\quad\quad\quad\quad\quad\quad\quad\quad\quad\quad\quad\quad]{
	            \label{fig:smooth_con}
	            \includegraphics[width=.4\linewidth]{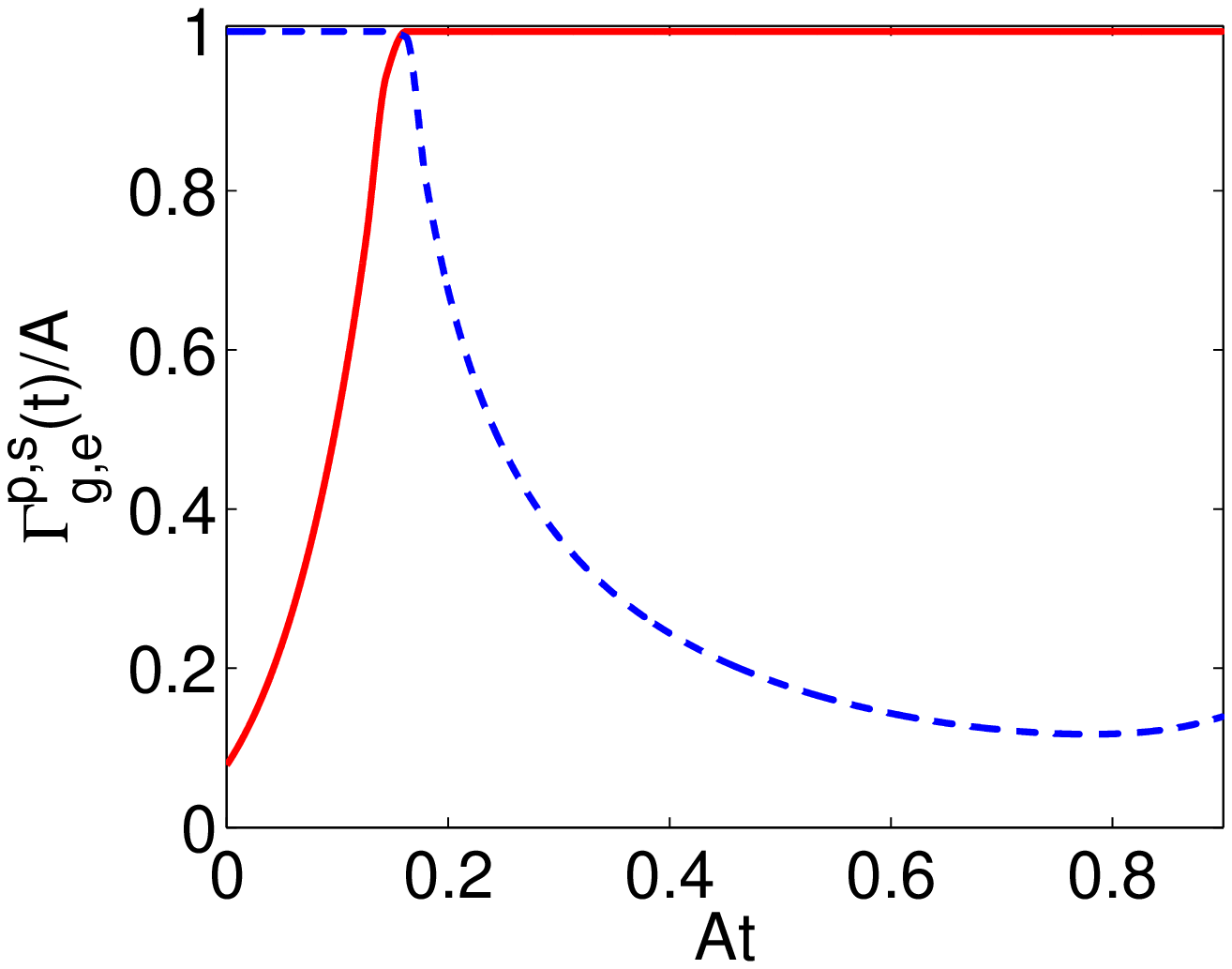}} &
      \subfigure[\quad\quad\quad\quad\quad\quad\quad\quad\quad\quad\quad\quad\quad\quad\quad\quad]{
	            \label{fig:smooth_pop}
	            \includegraphics[width=.4\linewidth]{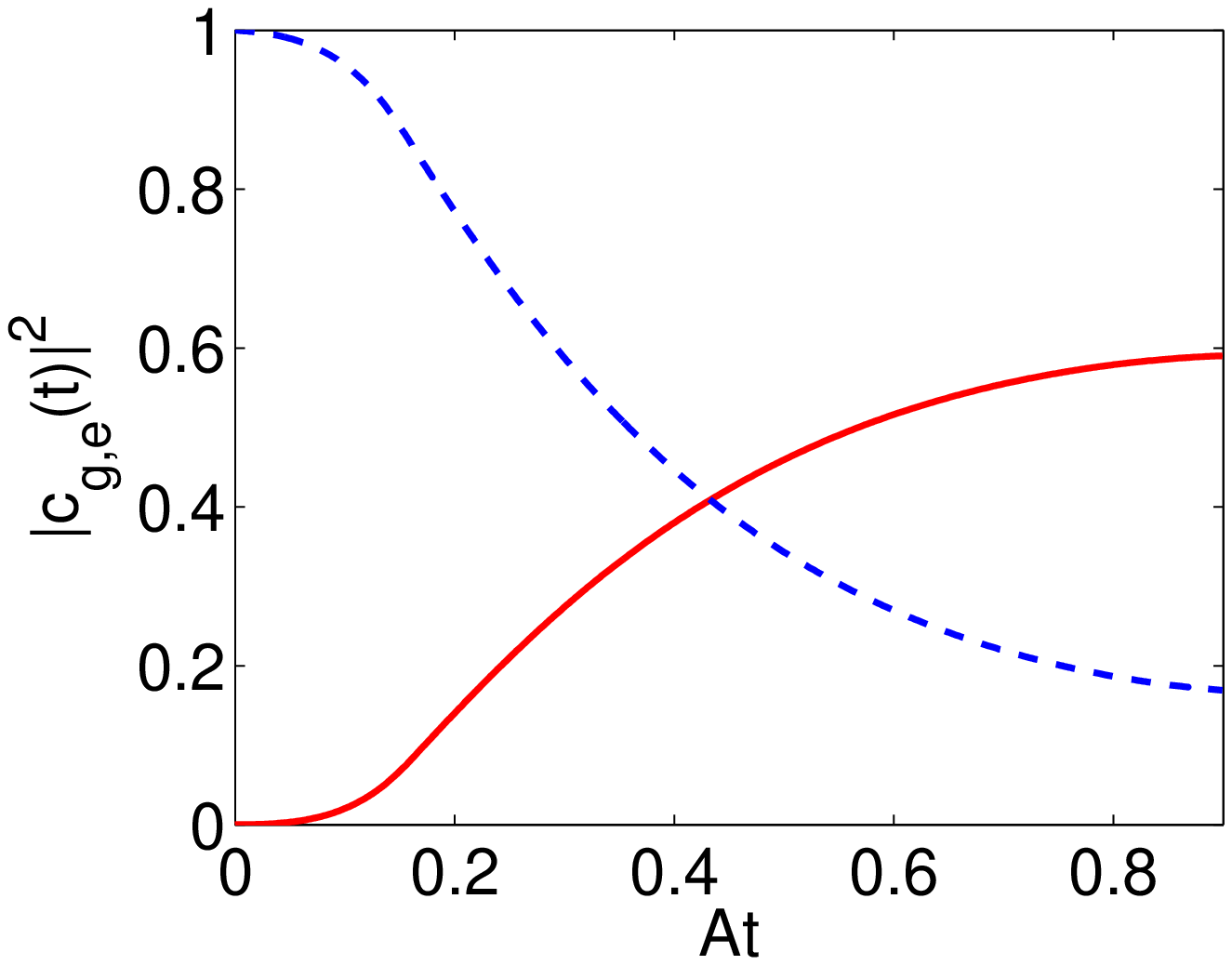}}
		\end{tabular}
\caption{(Color online) (a) Optimal ionization pulses $\Gamma^p_g(t)$ (red solid line) and $\Gamma^s_e(t)$ (blue dashed line) for normalized duration $AT=3$, in the case of effective two-photon resonance $\delta=0$ and $R=1/4$. (b) Corresponding evolution of populations of the ground (blue dashed line) and excited states (red solid line). (c) Optimal ionization pulses obtained for normalized duration $AT=0.9$, $\delta\neq 0$ and $R=1/16$. (d) Corresponding evolution of populations. (e) Smoothed version of the pulses of Fig. \ref{fig:Con_opt_d}. (f) Corresponding evolution of populations.}
\label{fig:Ex2}
\end{figure*}

The different behavior of improvement for increasing $R$, when $\delta=0$ and $\delta\neq 0$, is better demonstrated in Fig. \ref{fig:Eff_R}. There, we display the maximum excited-state population obtained with the optimal pulses (in the limit of large $AT$) as a function of parameter $R$, with step $\delta R=0.05$, for the case of effective two-photon resonance $\delta=0$ (red circles) and nonzero $\delta$ from Eq. (\ref{d}) (red squares). The isolated points at the specific values $R=0, 1/16, 1/4, 1$ indicate again the best efficiencies obtained in Ref. \cite{Vitanov97} using Gaussian STIRAP pulses. Now it is clear that for $\delta=0$ the improvement becomes better with increasing $R$, while for $\delta\neq 0$ becomes worse. This behavior can be explained if we recall that in general a larger parameter $R$ corresponds to a shorter effective duration available for the transfer. In the absence of two-photon detuning, the Gaussian STIRAP pulses can exploit the longer times which are available for smaller $R$ and obtain a transfer efficiency close to the optimal. But the presence of two-photon detuning deteriorates STIRAP, which cannot exploit the longer available durations for smaller $R$, to the same extent as the optimal method does.

In Fig. \ref{fig:Eff_vs_q} we plot the maximum excited-state population obtained with the optimal pulses (in the limit of large $AT$) as a function of Fano parameter $q$, with step $\delta q =0.5$, for the four different values of $R$ used throughout this paper. The isolated points at $q=-6$ indicate the best efficiencies obtained in Ref. \cite{Vitanov97} with Gaussian STIRAP pulses for the same values of $R$. Observe that the maximum efficiency increases with increasing $|q|$, except of course the case $\delta=0, R=0$, where equals unity for all $q$ (obtained in the limit of large $AT$). The improvement is more dramatic for the case with $\delta\neq 0$, shown in Fig. \ref{fig:Eff_q_full}. This can be understood if we recall that in section \ref{sec:linear} we identified $|q|/2$ as the frequency separation between the adiabatic eigenstates, thus its increase reduces drastically the deteriorative influence of the effective two-photon detuning.

In Fig. \ref{fig:Ex2} we display the optimal ionization widths and the corresponding evolution of populations for the two pairs of $\delta, R$ also used in Fig. \ref{fig:Ex1}. Specifically, we use the values $\delta=0, R=1/4, AT=3$ (first row) and $\delta\neq 0, R=1/16, AT=0.9$ (second row). The (finite) normalized durations for each case are selected such that the obtained efficiency closely approaches the maximum listed in the last row of Table \ref{tab:eff}. Observe that the optimal pulses have the bang-interior and interior-bang form described in Sec. \ref{optimal_solution}. And, although both pulses have nonzero values at the initial and final times, they follow in general the counterintuitive pulse order of STIRAP, since $\Gamma^p_g(t)$ (red solid line) corresponds to the pump pulse and $\Gamma^s_e(t)$ (blue dashed line) to the Stokes pulse. Note that the use of optimized pulses with nonzero Rabi frequency at the boundary times is not unusual in quantum control, and as a characteristic example we mention the Vitanov type optimized STIRAP pulses, see Ref. \cite{Clerk16} which describes their so-called ``superadiabatic" version. Another interesting observation is that in the second example, the control $\Gamma^p_g(t)$ maintains its maximum value for a longer portion of the available time interval than in the first example, compare the red solid lines in Figs. \ref{fig:Con_opt_d} and \ref{fig:Con_opt_d0}. The reason is that the incoherent terms (those involving $R$) in system Eqs. (\ref{state_system}) are proportional to $u_1^2\sim\Gamma^p_g(t)$, thus for smaller values of $R$, as in the second example, larger values of $\Gamma^p_g(t)$ can overall improve the efficiency despite the small increase of the incoherent losses. Additionally, from Eq. (\ref{d}) we see that larger values of $\Gamma^p_g(t)$ and thus of $u_1$ reduce the effective two-photon detuning, which is taken into account in the second example.

One of the striking differences between the optimal pulses and the Gaussian pulses is that they are not smooth but they have a kink, at the point where they change form from interior to bang or vice versa. In order to evaluate the implications of this feature, we find the transfer efficiency using a smoothed version of the pulses of Fig. \ref{fig:Con_opt_d}, displayed in Fig. \ref{fig:smooth_con}, which might be more suitable for practical implementation. The smoothed pulses are obtained by undersampling the original pulses by a factor of 20 and then using cubic interpolation for the sampled points. In Fig. \ref{fig:smooth_pop} we display the evolution of populations for the smoothed pulses, which looks identical to Fig. \ref{fig:Pop_opt_d}, obtained with the original optimal pulses. Actually, there is a very slight decrease in the transfer efficiency.

\begin{figure}[t]
 \centering
		\begin{tabular}{c}
     	
      \subfigure[\quad\quad\quad\quad\quad\quad\quad\quad\quad\quad\quad\quad\quad\quad\quad\quad]{
	            \label{fig:error1}
	            \includegraphics[width=.85\linewidth]{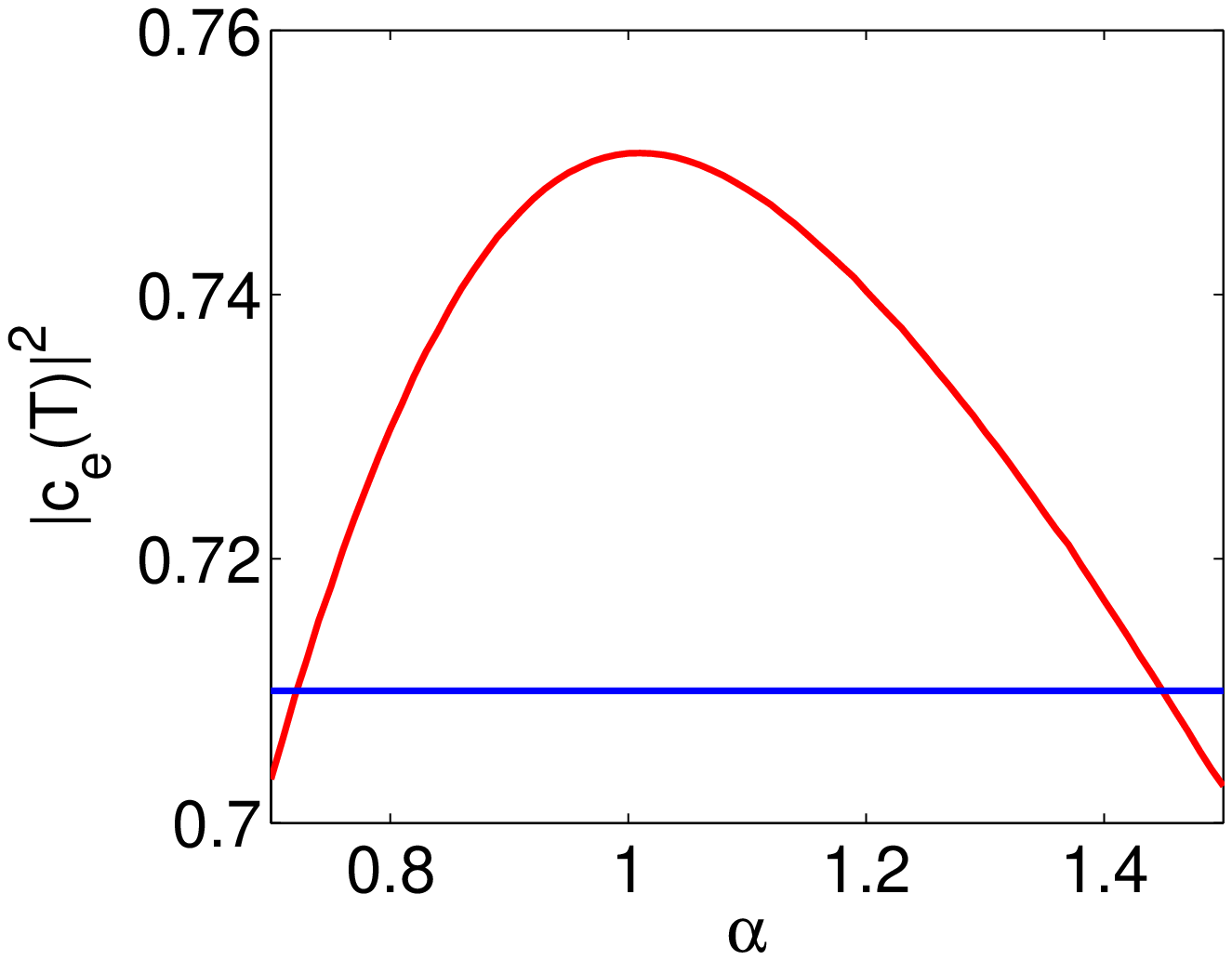}} \\
      \subfigure[\quad\quad\quad\quad\quad\quad\quad\quad\quad\quad\quad\quad\quad\quad\quad\quad]{
	            \label{fig:error2}
	            \includegraphics[width=.85\linewidth]{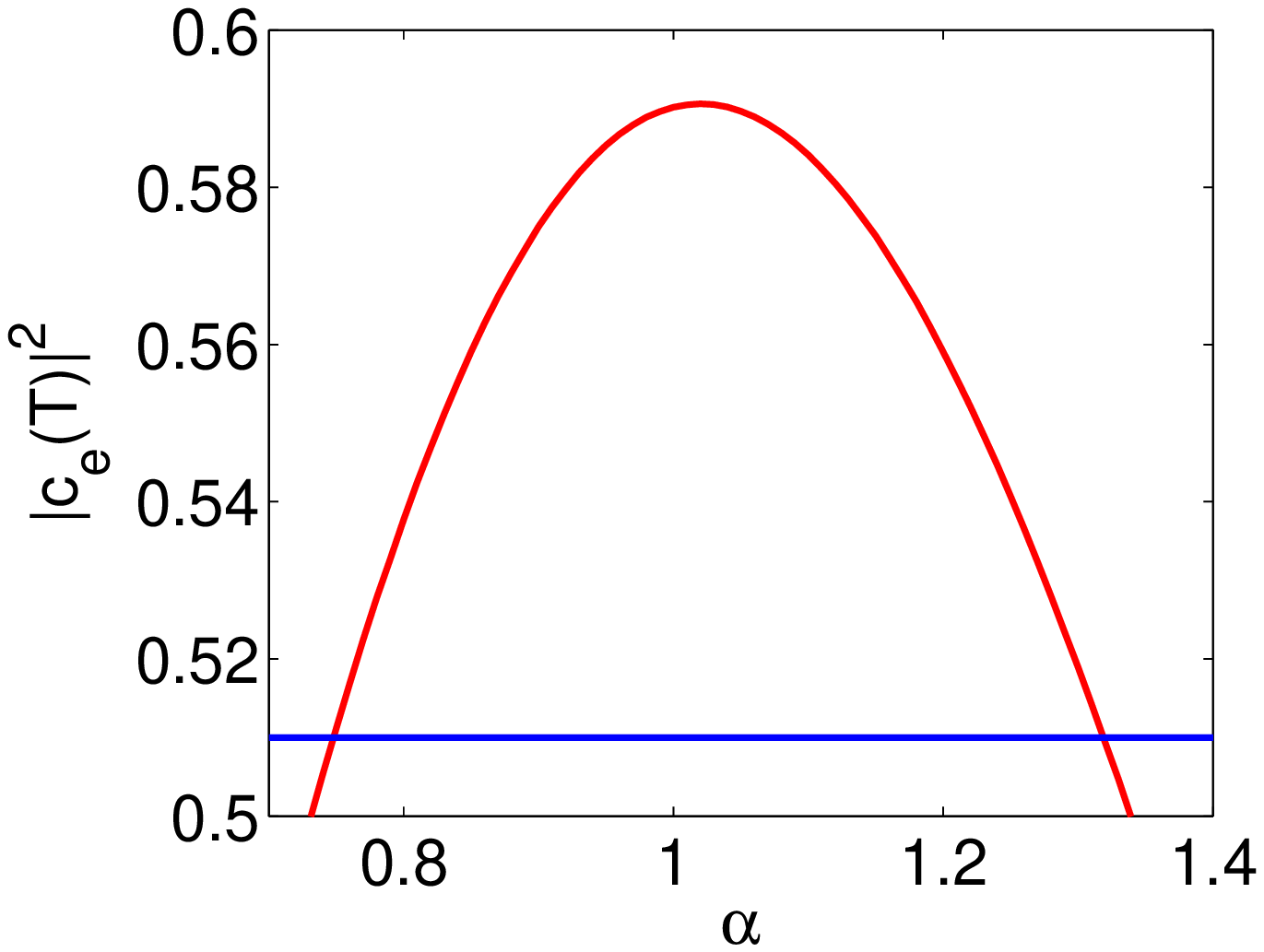}}
		\end{tabular}
\caption{(Color online) Excited state population obtained with the distorted pulses $\alpha\Gamma_{g,e}^{p,s}(t)$, using as reference pulses ($\alpha=1$) (a) the pulses shown in Fig. \ref{fig:Con_opt_d0}, (b) the pulses shown in Fig. \ref{fig:Con_opt_d}.}
\label{fig:robustness}
\end{figure}

We close by investigating the robustness of the proposed method. We consider the distorted pulses $\alpha\Gamma_{g,e}^{p,s}(t)$, where $\alpha$ is the distortion parameter. Note that, since $\alpha$ eventually multiplies the right hand sides of system equations, see Eqs. (\ref{state_system}) and (\ref{d}), it can also be used to rescale time as $t'=\alpha t$, thus $\alpha>1$ corresponds also to pulse dilation, while $\alpha<1$ to pulse contraction. In Figs. \ref{fig:error1}, \ref{fig:error2} we display with red solid lines the transfer efficiency obtained when the distorted pulses are applied to the system, corresponding to the examples shown in the first and second row of Fig. \ref{fig:Ex2}, respectively. The horizontal blue lines indicate the best efficiency obtained with Gaussian pulses, \emph{without} taking into account any error. Observe that the advantage of our method over the undisturbed Gaussian pulses is maintained for an appreciable range of the distortion parameter. The noticed asymmetry, where the performance is better for $\alpha>1$, is because these $\alpha$ values correspond to larger pulse areas.

\section{Conclusion}

\label{sec:conclusion}

We used optimal control theory to find pulses which maximize the population transfer between two bound states coupled via a continuum of states. We obtained better efficiencies than with the standard Gaussian STIRAP pulses, while the degree of improvement depends on whether we take into account the effective two-photon detuning, as well as the size of incoherent ionization. The present work is expected to be useful for applications involving population transfer between bound states through a continuum, for example coherence effects, like population trapping and electromagnetically induced transparency, optical analogs for light waves propagating in waveguide-based photonic structures, and qubits coupled via a continuum of bosonic or waveguide modes.


\end{document}